\documentclass[a4paper,fleqn]{cas-sc}
\usepackage{amsmath,amsfonts}
\usepackage[authoryear]{natbib}
\usepackage{subfigure}
\usepackage{threeparttable}
\usepackage{float}
\usepackage{multirow}
\usepackage{booktabs}
\usepackage{setspace}
{}
{}
{}
\usepackage{hyperref}

\begin{document}
\let\WriteBookmarks\relax
\def\floatpagepagefraction{1}
\def\floatpagepagefraction{2}
\shorttitle{IPF-HMGNN: A novel integrative prediction framework for metro passenger flow}
\shortauthors{Lu}

\title [mode = title]{IPF-HMGNN: A novel integrative prediction framework for metro passenger flow}                    

\address[1]{School of Transportation, Southeast University, Nanjing 211189, China}
\address[2]{Department of Civil Engineering, Monash University, Melbourne, VIC 3800, Australia}
\address[3]{College of Transportation Engineering, Chang’an University, Xi’an 710064, China}
\address[4]{School of Traffic and Transportation, Beijing Jiaotong University, Beijing 100091, China}
\author[1,2]{Wenbo Lu}[style=chinese]
\author[1]{Yong Zhang}[style=chinese]
\author[2]{Hai L.Vu}
\cormark[1]
\author[3]{Jinhua Xu}[style=chinese]
\author[4]{Peikun Li}[style=chinese]

\cortext[cor1]{Department of Civil Engineering, Monash University, Melbourne, VIC 3800, Australia}
\cortext[cor1]{E-mail: Hai.Vu@monash.edu}

\begin{abstract}
The operation and management of the metro system in urban areas rely on accurate predictions of future passenger flow. While using all the available information can potentially improve on the accuracy of the flow prediction, there has been little attention to the hierarchical relationship between the type of tickets collected from the passengers entering/exiting a station and its resulting passenger flow. To this end, we propose a novel Integrative Prediction Framework with the Hierarchical Message-Passing Graph Neural Network (IPF-HMGNN). The proposed framework consists of three components: initial prediction, task judgment and hierarchical coordination modules. Using the Wuxi, China metro network as an example, we study two prediction approaches (i) traditional prediction approach where the model directly predicts passenger flow at the station, and (ii) hierarchical prediction approach where the prediction of ticket type and station passenger flow are performed simultaneously considering the hierarchical constraints (i.e., the sum of predicted passenger flow per ticket type equals the predicted station aggregated passenger flow). Experimental results indicate that in the traditional prediction approach, our IPF-HMGNN can significantly reduce the mean absolute error (MAE) and root mean square error (RMSE) of the GNN prediction model by 49.56\% and 53.88\%, respectively. In the hierarchical prediction approach, IPF-HMGNN can achieve a maximum reduction of 35.32\% in MAE and 36.18\% in RMSE, while satisfying the hierarchical constraint.
\end{abstract}
\maketitle

\section{Introduction} \label{1}
The metro system, due to its economic, punctual, and efficient characteristics, is one of the most important and popular  choices for travel in cities worldwide. The rapid growth in passenger demand due to the increased urbanization poses great challenges to the operation and management of the metro system in urban areas. Accurate and real-time short-term passenger flow (STPF) prediction is thus essential for decision making in the operational management of the metro system (\citealp{wang2021metro}; \citealp{lu2023mul}). Deep learning (DL) models are widely used in metro STPF prediction due to its powerful nonlinear fitting abilities. Researchers have developed various DL prediction models from the perspective of time and space modelings utilizing temporal and spatial characteristics, respectively. While the time-series models are difficult to model spatial correlations, Graph Neural Network (GNN) models with spatial graph structures have been used frequently in many applications including short term prediction (\citealp{xu2023multi}; \citealp{yang2023estimating}). GNN models use message passing functions to aggregate information from neighboring nodes in a graph topology representing a physical network which enables the GNN prediction model to capture spatial correlation between stations while retaining the temporal features of the data. Furthermore, as it is difficult for a single graph to capture the complex dependencies, GNN models based on multi-view modeling have been popular in the research community for STPF research (\citealp{jin2021hetgat}; \citealp{lu2021dual}; \citealp{wu2023learning}).

While many GNN prediction models have been developed for metro station passenger flow prediction (\citealp{8536904}; \citealp{ZHANG2022103659}; \citealp{FANG2024122550}), several gaps remained to be addressed including: 1) passenger flow prediction relied solely on input features of historical passenger flow obtained at individual time points without considering its local or global trend; 2) little research on prediction of passenger flow per ticket type; 3) lacking consideration of hierarchical constraints where the sum of prediction value for sub passenger flow per ticket type must be equal to the predicted aggregate station passenger flow.

To address the above gaps, in this paper we propose to transform the historical passenger flow sequences from a point to patch input via aggregating 15-minute interval data into patches over a longer period of time (e.g., 60 minutes). This allows the learning of local and global features embedded in each data patch consisting a number of data points. 

Furthermore, in order to capture the hierarchical relationships between the sub passenger flow per ticket type and the station passenger flow, we introduce a novel hierarchical message-passing mechanism in a GNN model. 
We also develop a hierarchical coordination model to solve the hierarchical constraint problem which can be directly integrated into the above GNN prediction model using a deep learning fusion or concatenation mechanism (\citealp{FANG2024122550}; \citealp{bao2022forecasting}). 

In summary, the main contributions of the paper are listed below.
\begin{itemize}
\item develop a new input data segmentation method with patching operation and depthwise separable convolution to extract global and local features of different patch blocks;
\item establish a general hierarchical message-passing mechanism in GNN model for metro STPF prediction problem. The model considers the potential hierarchical relationships in the predicted passenger flows;
\item construct a DL-based hierarchical prediction (HP) coordination module and facilitate seamless integration into existing DL prediction models through the utilization of DL’s concatenation mechanism;
\item propose a novel Integrative Prediction Framework with the Hierarchical Message-Passing Graph Neural Network (IPF-HMGNN) for both the HP and traditional prediction tasks with or without hierarchical constraints, respectively; and
\item demonstrate the utility of the proposed IPF-HMGNN framework in the STPF prediction problems using real data collected from the Wuxi, China metro network.
\end{itemize}

The rest of the paper is organized as follows. Section 2 reviews and discusses the existing research on metro STPF. In the Section 3, we introduce the model notations, problem formulation and the concept of GNN and HP approaches and describe the details of the IPF-HMGNN framework in Section 4. Next, the IPF-HMGNN framework is applied for the Wuxi,  China metro network, and the results are discussed in Section 5. Finally, the conclusion is presented in Section 6.
\section{Literature Review}
The central theme of this study was the input feature extraction and multi-level GNN modeling for the STPF problem. Below, we will review existing research on input data processing, hierarchical relationship construction, and GNN modeling and identify the knowledge gaps.
\subsection{Decomposition and aggregation modeling}
To perform STPF prediction, the decomposition and aggregation techniques have generally been adopted for input data processing (\citealp{li2023traffic}; \citealp{tang2022seasonal}). At the same time, these techniques can also generate potential passenger flow hierarchical relationships. Decomposition methods decomposed passenger flow sequences into sub-sequences at different scales. Then, the predictive models were built using these sub-sequences data. The potential hierarchical relationship can be formed between the sub-sequences and original passenger flow sequence, such as station passenger flow and its sub-sequences
of different scales. On the other hand, the data aggregation approach primarily involved categorizing or clustering multiple passenger flow sequences. Passenger flow sequences belonging to the same cluster used the same model parameters. The potential hierarchical relationship can then be formed between the original multiple passenger flow and their clusters, like the passenger flow of stations with the same functional type and the total passenger flow of this cluster. Herein, we review the existing research on using decomposition and aggregation techniques for STPF problems.

There have been several methods such as Empirical Mode Decomposition (EMD), Variational Mode Decomposition (VMD), and Seasonal and Trend decomposition
using Loess (STL) proposed to predict metro passenger flow (\citealp{huang2023deaseq2seq}; \citealp{liu2020short}; \citealp{wei2012forecasting}; \citealp{zhang2020lightgbm}). The first two methods decomposed passenger flow data into components representing features at various time scales or frequency ranges while the third method dissected the passenger flow sequence into trend, seasonality, and residuals. The decomposition outcomes were more lucid and accessible to comprehend. However, it assumed that the time series was stationary and thus unable to handle nonlinearity, non-stationarity, or complex data (\citealp{chen2020forecasting}; \citealp{qin2019effective}).

In contrast, the aggregation method facilitated information aggregation among similar passenger flow sequences from different stations (\citealp{lu2023mohp}; \citealp{sajanraj2021passenger}) where clustering algorithms were used to learn feature embeddings for
similar types of stations. For instance, \cite{wei2022cluster} employed clustering algorithms to partition all metro stations
into several groups where stations belonging to each group used the same feature embeddings. In another study,  \cite{tu2022forecasting} suggested that the functional type of metro stations (e.g., residential station or transportation hub) can be determined by their surrounding Points of Interest (POI). The passenger flow sequences of stations with the same functional type were similar which can then facilitate learning and predicting tasks (\citealp{LIU2020102561}; \citealp{YU2022103299}).
\subsection{Deep learning methods based on GNN model}
Metro networks can be treated as the graph-structured data, necessitating the presence of a global topology and station relationships to improve the STPF prediction accuracy. GNN models are suitable for this purpose, as they can establish connections among stations and effectively leverage them for prediction purposes. Specifically, for every station, GNN can extract and propagate information of the neighbor via message passing functions, which enables nodes to gather information from their neighbors and update their own features accordingly. The trained GNN model thus utilized the dependence between passenger flows observed from connected stations.

Constructing graphs and mining relationships between stations was critical in utilizing GNN models for STPF prediction tasks. The physical metro network was a basis for constructing a typical graph where the relationship between stations is established based on their physical topological connections. For example, \cite{wang2021metro} constructed a hypergraph based on the metro physical topology to capture the higher-order relationships between stations. However, this approach may not adequately reflect the actual spatial dependencies between stations and recent studies have proposed alternative methods to improve the GNN graph structure. For instance,  \cite{zhao2023adaptive} used a trainable adaptive adjacency matrix to obtain the graph structure for STPF.

Due to the complex and diverse spatial correlations between stations, GNN models of multi-graph fusion have recently been adopted. In particular, \cite{liu2020physical} constructed graphs from physical connections of stations, passenger flow sequence similarities, and functional correlations respectively to learn effective spatiotemporal representations between stations. \cite{bao2022forecasting} considered the geographical distance, functional similarity, and demand pattern to construct the GNN graphs. Recently, \cite{li2023ig} developed connectivity, similarity, and temporal correlation graphs to model station interactions; \cite{CHEN2023} processed physical topology graphs and origin-destination (OD) graphs to capture spatial features;  \cite{wang2023network} explored adjacency similarity, geographical location similarity, and trend similarity. Moreover, \cite{xie2023spatio} employed a representation model with adaptive and spatial structure awareness to learn the dynamic spatial dependence between stations without a predefined graph adjacency matrix.

The above studies primarily focused on the construction of the spatial correlation graphs, graph models, and multi-view graphs fusion. However, they only considered a single-level passenger flow modeling and without multi-level hierarchy information transmission and fusion. For instance, when predicting station passenger flow, only the spatial and temporal correlations between stations were modeled without considering the inter-layer relationship between sub passenger flow per ticket type and total flow at a station. Table~\ref{tab} summarizes the previous studies on input features, hierarchical structure, and prediction task and highlight the knowledge gaps in the state-of-the art STPF research.
\begin{table}[ht]
  \centering
  \setlength{\tabcolsep}{1.2mm}
  \caption{A comparison of some existing metro SPTF prediction models using GNN.}
    \begin{tabular}{cccccc}
    \toprule
    Authors (Year)	&Input feature&	Hierarchical Structure & Constructed graph  &	Research object	& Predicted Task \\
    \midrule
    \cite{liu2020physical} &Points	&-	& MG&Station	&TP\\
   \cite{bao2022forecasting}	&Points	 &-	&MG &Station	&TP\\
   \cite{zhao2023adaptive}	&Points	 &-	&AG&Station&	TP\\
   \cite{li2023ig}	&Points	 & -	&MG&Station	&TP\\
   \cite{CHEN2023}	    &Points  &	-	&MG&Station&	TP\\
   \cite{wang2023network}	&Points	  &-	&MG&Station&	TP\\
   \cite{xie2023spatio}	&Points	  &-	&AG&Station&	TP\\
   Our proposed    &Patches  & \checkmark	& SG and MG&Station and Tickets &	TP and HP\\
    \bottomrule
    \end{tabular}
    \begin{tablenotes}
        \footnotesize
        \item[1] MG: multi-view graphs; AG: adaptive graph; SG: single-view graph; TP: traditional prediction; HP: hierarchical prediction.
      \end{tablenotes}
  \label{tab}
\end{table}
\section{Preliminaries}
\subsection{Problem formulation} \label{3.1}
\noindent \textbf{Definition 1 (Graph):} In this paper, a general network representation is denoted as a undirected graph $G = (V, E)$, where $V, E$ are the set of nodes and edges, respectively. 

\noindent \textbf{Definition 2 (Multi-layer Graph):} We define a multi-layer graph consisting of $m+1$ graphs stacked horizontally. Specifically, the undirected graph on the $m^{th}$ layer is denoted as $G_m = (V_m, E_m)$. 

\begin{figure}[ht]
	\centering
	\includegraphics[width=11cm]{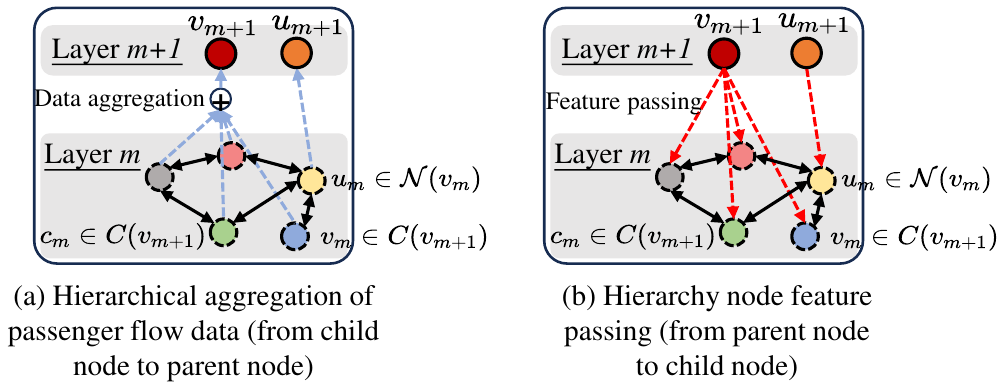}
	\caption{An example of a multi-layer graph.}
	\label{figure}
\end{figure}

\noindent \textbf{Definition 3 (Hierarchical Node):} Let $C(v_{m+1})$ be a set of nodes in the $m^{th}$ layer that have been grouped together using a clustering method according to their characteristics (e.g. functional similarity or demand pattern), in case of nodes in a graph representing the metro stations where the node $v_{m+1}$ (shown in Fig.~\ref{figure}) is in the $(m+1)^{th}$ layer and referred to as a parent node having an aggregated characteristic of all the nodes in the set $C(v_{m+1})$. The $c_m, v_m \in C({v_{m + 1}})$ nodes are referred to as the child node of $v_{m+1}$. Let ${\cal N}(v_m)$ be the set of nodes in the $m^{th}$ layer that are connected with node $v_m$, thus $u_m \in {\cal N}(v_m)$ is connected to $v_m$ albeit may belong to a different parent node.

\noindent \textbf{Definition 4 (Vertical Graph):} We define a vertical graph $G_{m+1}^v = (V_{m+1}^v, E_{m+1}^v)$ consisting of node $v_{m+1}$ in the $(m+1)^{th}$ layer and its child nodes set $C({v_{m + 1}})$ in the $m^{th}$ layer where the node $v_{m+1}$ points to its child nodes as shown in Fig.~\ref{figure}. 

\noindent \textbf{Definition 5 (Metro Passenger Flow):} in a graph $G_m = (V_m, E_m)$ at $m^{th}$ layer representing the physical metro network, the passenger flow of node $v_m$ at time $t$ is defined as ${x^t}({v_m})$. The vector of passenger flows of node $v_m$ over the $L$ time periods up to time $t$ is denoted as ${x^{t - L:t}}({v_m}) = [{x^{t - L}}({v_m}),...,{x^{t - 1}}({v_m}),{x^t}({v_m})] \in {{\bf{R}}^{1 \times L}}$. The matrix of passenger flows of the graph $G_m$ is denoted as ${X^{t - L:t}}({V_m})$.

\noindent \textbf{Definition 6 (Hierarchical Passenger Flow):} The passenger flow of node $v_{m+1}$ at time $t$ is obtained by summing all the passenger flows of its child nodes, i.e., $x^t({v_{m+1}})=\sum {\left \{x^t({c_m}), x^t({v_m})\right \}}, {\rm{\forall }}{v_m, c_m} \in C({v_{m + 1}})$, as shown in Fig.~\ref{figure}.(a). 

\noindent \textbf{Definition 7 (Hierarchical information passing):} The hierarchical information such as node's embedded feature is transferred from parent nodes to child nodes. For example, node $v_m$ obtains the information of parent node $v_{m+1}$ in $(m+1)^{th}$ layer, as shown in Fig.~\ref{figure}.(b). 

\noindent \textbf{Definition 8 (Prediction layer):} We define the layer where the nodes providing the prediction results are located as the prediction layer. Table~\ref{tab1} summarizes the mathematical notations used in this paper.

\begin{table}[ht]
  \centering
  \caption{Summary of mathematical notation.}
    \begin{tabular}{cc}
    \toprule
    Notation&	Description\\
    \midrule
    $G,E,V$ & Graph, edges, and node sets (also referred to as the total number of nodes)\\
    $v$ & General representation of nodes \\
    ${\cal N}({v})$&Set of nodes $u$ directly connected to node $v$, $u \in {\cal N}({v})$\\
    ${G_m},{E_m},V(m),V_m$& $m^{th}$ layer graph, edges, node set, and node count\\
    $V(Pr),{V_{Pr}}$&Prediction layer node set and node count\\
    ${v_m}$&Nodes in the $m^{th}$ layer, $v \in V_m$\\
    ${\cal N}({v_m})$&$m^{th}$ layer connected nodes set of $v_{m}$\\
    ${u_m}$&Connected node ${u_m}$ for node ${v_m}$, $u_m \in {\cal N}({v_m})$\\
    $C({v_{m + 1}})$&$m^{th}$ layer child nodes set of $v_{m + 1}$\\
    ${c_m}$& ${c_m}$ has the same parent node as node ${v_m}$, $c_m$ and ${v_m} \in C({v_{m + 1}})$ \\
    $L$&Lookback window length\\
    $T$&Prediction length\\
    ${x^t}({v_m})$&Passenger flow of node ${v_m}$ at time $t$\\
    ${x^{t - L:t}}(v),{x^{t - L:t}}({v_m})$&Passenger flow vector for nodes ${v}$ and ${v_m}$ during the time period $t-L$ to $t$ \\
    ${X^{t - L:t}}$& Passenger flow matrix during the time period  $t-L$ to $t$\\ 
    ${X^{t - L:t}}({V_m})$& Passenger flow matrix of the $m^{th}$ layer during the time period $t-L$ to $t$\\
    \bottomrule
    \end{tabular}
  \label{tab1}
\end{table}

\noindent \textbf{Problem formulation:} Given the input passenger flow matrix ${X^{t - L:t}}$ during the time period $L$ up to time $t$, the objective is to establish a fitting model $f( \cdot )$ to minimize the error $Loss({\widetilde X^{t + 1:t + T}},{X^{t + 1:t + T}})$  between the predicted values ${\widetilde X^{t + 1:t + T}} = f({X^{t - L:t}})$ for a future time period $T$ and the observed values ${X^{t + 1:t + T}}$, where the $Loss$ is the error function. 
\subsection{Message Passing in GNN}
The message-passing of GNN allows nodes to exchange information through the connected structure of the graph. 
Assuming that node $u$ is connected with $v$, the message passing consists of the following steps:

\noindent \textbf{(1) Aggregating Information (Aggregation):} For node $v$ at any time $t$, the information from its connected nodes is aggregated using operations such as summation or averaging:
\begin{align}
{m_v} = \rho \left( {\left\{ {{{\bf{h}}_u},{\rm{\forall }}u \in {\cal N}(v)} \right\}} \right)
\end{align}

where the ${m_v}$ is message vector of node $v$; $\bf{h}_u$ is the node feature for $u$; $\rho$ is the aggregation function.

\noindent \textbf{(2) Updating Node Representation (Update):} the aggregated information is combined with the current node's feature using an update function $\phi$ to generate the new representation of node $v$ at any time $t$:
\begin{align}
{\bf{h}}_v^\prime  = \phi \left( {{{\bf{h}}_v},{m_v}} \right)
\end{align}

where the $\bf{h}_v$ is the current node feature of $v$; It should be noted that each node only uses the current feature of the connected node, not the updated node feature.
\subsection{Hierarchical prediction problem}
\begin{figure}[ht]
	\centering
	\includegraphics[width=9cm]{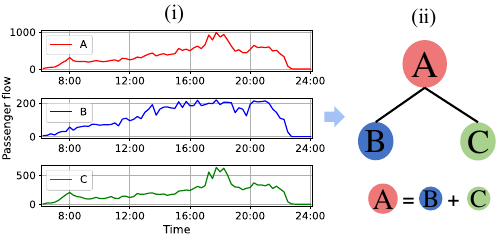}
	\caption{The hierarchical structure existing in the metro passenger flow sequence.}
	\label{figure1}
\end{figure}

In the metro systems, the hierarchical constraints should be imposed when predicting simultaneously the passenger flow per ticket type
and the total flow at a station. For instance, the sum of predicted passenger flows per ticket type must be equal to the predicted aggregated station passenger flow, which refers to as a hierarchical prediction (HP) task. The traditional HP approach include top-down (TD), bottom-up (BU), and middle-out (MO) method. These methods involve obtaining the prediction results from the bottom, top, and middle layers, respectively. Subsequently, predictions for other layers are obtained by incrementally moving up or down and decomposing according to the proportion of the child node in the parent node. In contrast, the coordination methods obtain the prediction results for all layers simultaneously. Then, the calibration matrix $P$ and hierarchy relationship matrix $H^r$ are used to coordinate all predicted values to satisfy the hierarchical constraints \citep{ANDERER20221405}. The process of the coordination method can be represented as: 
\begin{align}
\tilde y = H^rP \hat y
\label{eq3}
\end{align}

 where $\hat y$ and $\tilde y$ are the initial and coordinated prediction values, respectively. For example, let us consider three nodes $A$, $B$ and $C$ in Fig.~\ref{figure1} where their passenger flows satisfies $A=B+C$. Among them, $A$ represents an individual station, and $B$ and $C$ represent the different ticket types at this station. Thus, $A$ is the parent node of $B$ and $C$. The hierarchy relationship matrix $H^r \in {{\bf{R}}^{3 \times 2}}$ is:
\begin{align*}
H^r = \left[\begin{array}{ll}
1 & 1  \\
1 & 0  \\
0 & 1
\end{array}\right] 
\end{align*} 

where the first row vector $[1,1]$ shows node $A$ is a parent node of both nodes $B$ and $C$ at the bottom layer, while the second $[1,0]$ and third $[0,1]$ row vectors indicate node $B$ and $C$ are their own parent node but not that of other. The calibrated matrix $P \in {{\bf{R}}^{2 \times 3}}$ is defined as
\begin{align*}
P  = \left[\begin{array}{lllllll}
p_{11} & p_{12} & p_{13} \\
p_{21} & p_{22} & p_{23} 
\end{array}\right]
\end{align*}

Denote the initial values of $A$, $B$, and $C$ in period $t$ as $\hat{A}_{t}$, $\hat{B}_{t}$, and $\hat{C}_{t}$ respectively, the process of coordination methods is as follows:
\begin{align}
\tilde{y}_t&=\left[\begin{array}{ll}
1 & 1  \\
1 & 0  \\
0 & 1
\end{array}\right] \cdot\left[\begin{array}{lll}
p_{11} & p_{12} & p_{13}\\
p_{21} & p_{22} & p_{23} 
\end{array}\right] \cdot \left[\begin{array}{l}
\hat{A}_{t} \\
\hat{B}_{t} \\
\hat{C}_{t}
\end{array}\right] \nonumber \\
&=\left[\begin{array}{lll}
\sum_{i=1}^{2} p_{i 1} & \sum_{i=1}^{2} p_{i 2} & \sum_{i=1}^{2} p_{i 3} \\
p_{11} & p_{12} & p_{13}  \\
p_{21} & p_{22} & p_{23} 
\end{array}\right] \cdot \left[\begin{array}{l}
\hat{A}_{t} \\
\hat{B}_{t} \\
\hat{C}_{t}
\end{array}\right] \nonumber \\
&=\left[\begin{array}{llll}
\sum_{i=1}^{2} p_{i1}\cdot\hat{A}_{t}
 + \sum_{i=1}^{2} p_{i2}\cdot\hat{B}_{t}
 +\sum_{i=1}^{2} p_{i3} \cdot\hat{C}_{t}\\
p_{11}\cdot\hat{A}_{t}
+ p_{12} \cdot\hat{B}_{t}
+ p_{13}\cdot\hat{C}_{t}\\
p_{21} \cdot\hat{A}_{t}
+ p_{22} \cdot\hat{B}_{t}
+ p_{23} \cdot\ \hat{C}_{t}
\end{array}\right]  \nonumber \\
& =\left[\begin{array}{llll}
\tilde{B}_{t}+\tilde{C}_{t}\\
\tilde{B}_{t} \\
\tilde{C}_{t}
\end{array}\right]
\end{align}

In order to obtain the matrix $P$, existing research includes using the generalized least squares \citep{hyndman2011optimal} and the minimum trace estimator \citep{wickramasuriya2018optimal}.
\section{Methodology}
\subsection{IPF-HMGNN Framework}
\begin{figure}[ht]
	\centering
	\includegraphics[width=14cm]{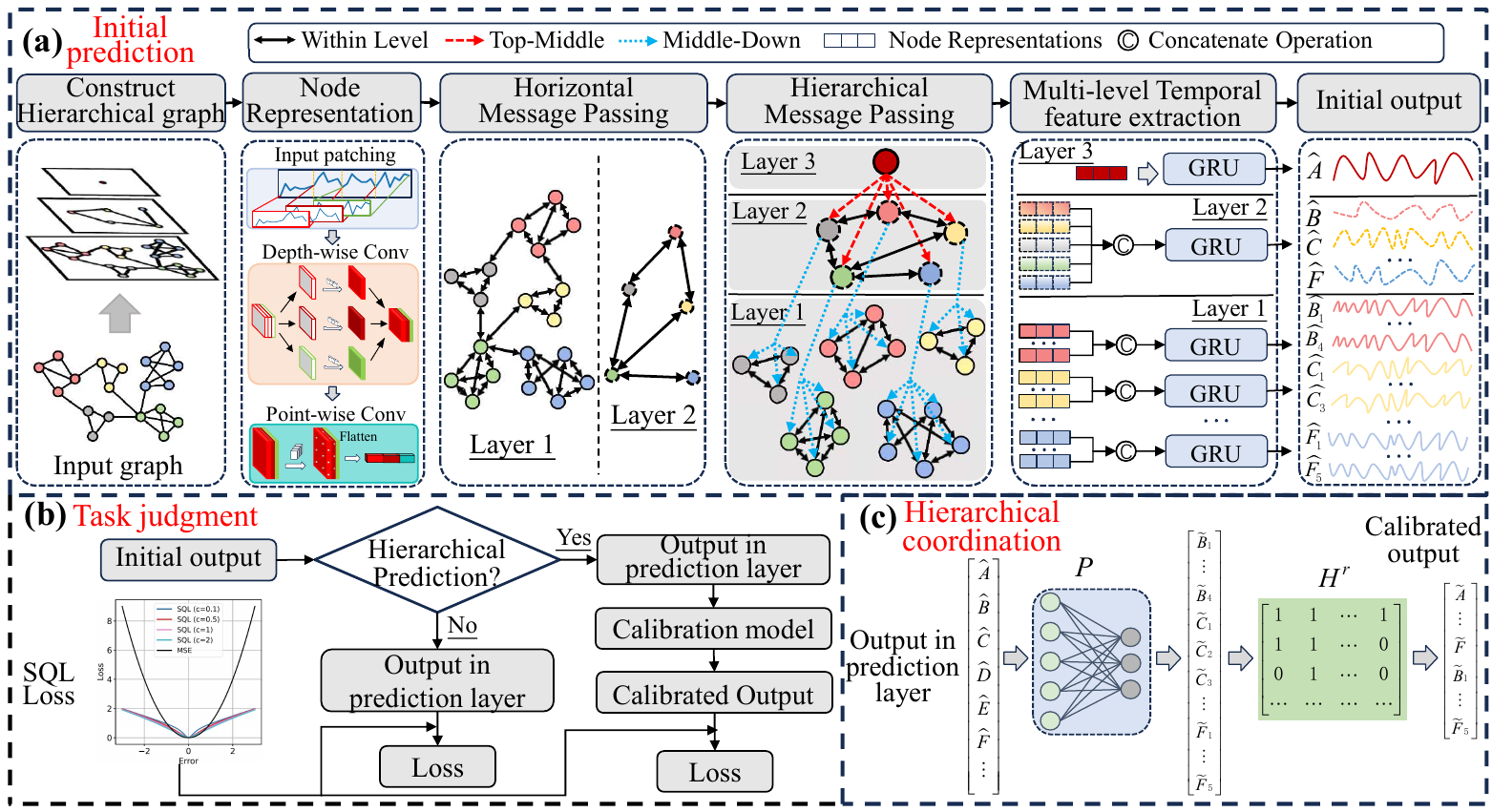}
	\caption{The overview of IPF-HMGNN framework; (a) the process for getting the initial output; (b) the task judgment module; (c) the process of prediction coordination. Note: The nodes in the input graph depend on the prediction task, which may be stations or ticket types; "Output in prediction layer" means that we only output the results of those nodes in the prediction layer. For example, we do not output the results of the middle layer (e.g. clustering layer) nodes. Specific examples can be found in Section \ref{5.3}.}
	\label{lu2}
\end{figure}

We propose below the IPF-HMGNN framework shown in Fig.~\ref{lu2} to address the knowledge gaps mentioned in Sec.\ref{1}, including: (1) exploring the local and global features within the passenger flow time series (2) achieving predictions of per ticket types passenger flow and (3) satisfying the prediction hierarchical constraints between prediction targets. To this end, the proposed framework includes three modules: initial prediction, task judgment and hierarchical coordination modules. 
The input of the IPF-HMGNN framework is the historical passenger flow sequences and multi-layer graphs, and the output is the prediction results of all nodes in the prediction layer. First, the initial prediction results are obtained, and then the type of prediction task (traditional prediction, TP or hierarchical prediction, HP) is determined in the task judgment part where the loss value is calculated. Hierarchical coordination is a part of the task judgment and is used to coordinate the prediction results when the prediction task is HP. To be specific, multiple layer graphs are constructed in the initial prediction module, which correspond to different tasks, such as ticket type and station passenger flow prediction. Wile constructing the multi-layer graphs, we  develop a new input data segmentation method with patching operation and depthwise separable convolution to extract global and local features of different patch blocks as the embedded feature of nodes in the different layers of the constructed multi-layer graph. We then propose a new GNN with hierarchical message passing mechanism to capture the spatial correlations. Subsequently, the GRU is used to obtain initial prediction values for different prediction tasks in multiple layers. In the task judgment module, it is judged whether the prediction task is TP or HP, so that different methods can be adopted to obtain the final predicted passenger flow and calculate the loss value. In the hierarchical coordination module, a new hierarchical coordination method based on a deep learning model is proposed for the HP task to adjust the different prediction values to meet the hierarchical constraints. In addition, to mitigate the impact of outliers, a more robust smooth quadratic loss (SQL) function is used to calculate the error values. Finally, the mini-batch approach is developed to train the IPF-HMGNN framework.
\subsection{Construct Hierarchical graph} \label{4.2}
\begin{figure}[ht]
	\centering
	\includegraphics[width=12cm]{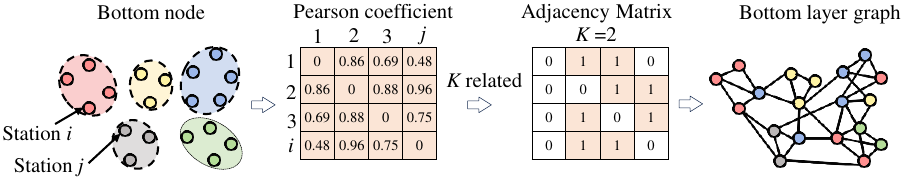}
	\caption{The process of constructing the bottom graph.}
	\label{lu4}
\end{figure}

To capture the potential relationship, we define the graph structures at different hierarchical levels by constructing a three-layer graph where the bottom layer is the first-layer graph (i.e., $m$=1) , the middle and top layers are the second- and third layer graphs (i.e., $m$=2, 3), respectively. It should be noted that the top-layer graph may contain only one node. The construction process of the bottom-layer graph is explained in Fig.~\ref{lu4} where the nodes represent the metro stations. Herein we calculate the passenger flow similarity between each station pairs in the studied metro network to construct a similarity matrix. The top $K$ station pairs with high similarity scores are then selected to construct an adjacency matrix. Specifically, for station $i$, its passenger flow time series is the average of the daily passenger flow statistics aggregated by time in the training set. Then, the similarity between stations $i$ and $j$ is measured by calculating the \textit{Pearson} coefficient \citep{lu2022measuring} of the passenger flow time series. Finally, the top $K$ most similar stations are selected to connect to station $i$.

\begin{figure}[ht]
	\centering
	\includegraphics[width=7.5cm]{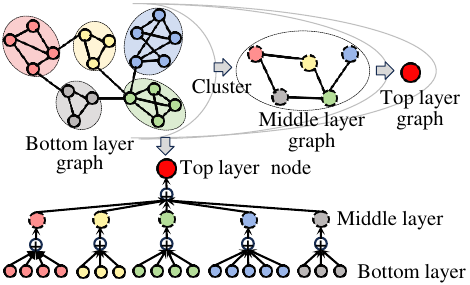}
	\caption{The process of constructing a hierarchical graph.}
	\label{lu3}
\end{figure}

Next, we employ a clustering algorithm (e.g. k-means \citep{FANG2024122550} in our numerical examples below) to capture the potential hierarchical relationships and construct the middle and top layer graph, as shown in Fig.~\ref{lu3}. In each cluster, the sum of passenger flows of the bottom-layer nodes belonging to the cluster constitutes the passenger flow of their parent node which is located in the middle-layer graph. Furthermore, all nodes in middle-layer graph are then aggregated into the top-layer graph node in a similar way or by summing the passenger flows from all the parent nodes in the middle layer.
\subsection{Input Patching Construct}
\begin{figure}[ht]
	\centering
	\includegraphics[width=9.0cm]{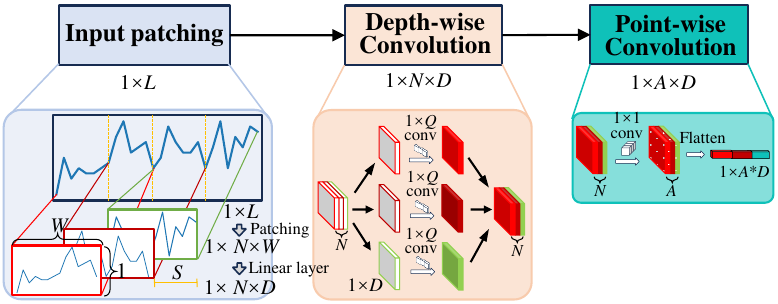}
	\caption{Input patching process.}
	\label{lu5}
\end{figure}

Next, we will introduce the process of extracting the global and local features of the input data. For nodes at different layers, we employ the patching operation to partition the input passenger flow sequence data into multiple blocks. As shown in Fig.~\ref{lu5}, for a node \(v_m\), a sliding window method with a window size of \(W\) and a step size of \(S\) are utilized to construct \(N\) passenger flow patches from the original input passenger flow sequence. Subsequently, a linear layer \citep{han2019predicting} is applied to obtain the embedding representation for each passenger flow patch. The process can be described as:
\begin{subequations}
\begin{align}
&{x^{emb}}({v_m}) = {\mathop{\rm Linear}\nolimits} ({x^{t - L:t}}({v_m})) \in {{\bf{R}}^{1 \times N \times D}} \\
&N=\left \lfloor \frac{L-W}{S}  \right \rfloor +1
\end{align}
\label{eq4}
\end{subequations}

where $\left \lfloor  \right \rfloor$ means the rounding down operation.

Furthermore, a convolutional neural network (CNN) is employed for extracting the local and global features of the different passenger flow patches. In previous studies, different-sized convolutional kernels are applied to convolve the single passenger flow patch, which makes it challenging to differentiate between local and global information. We utilize two convolutions to extract the local and global information, respectively. Specifically, a deep convolutional kernel with a size of $1 \times Q$ is employed to convolve within each passenger flow patch, achieving the local feature extraction. Subsequently, a  $A$-channel point-wise convolutional kernel is used to model the relationships between $N$ passenger flow patches for global feature extraction \citep{gong2023patchmixer}. The specific process is as follows:
\begin{subequations}
\begin{align}
&{x^d}({v_m}) = \sigma \{ {\rm{Depthwise(}}x^{emb}({v_m}))\}  \in {{\bf{R}}^{1 \times N \times D}} \\
&{x^p}({v_m}) = \sigma \{ {\rm{Pointwise(}}{x^d}({v_m}))\}  \in {{\bf{R}}^{1 \times A \times D}} 
\end{align}
\label{eq5}
\end{subequations}
\subsection{Multi-level Message Passing}
\begin{figure}[ht]
	\centering
	\includegraphics[width=6.5cm]{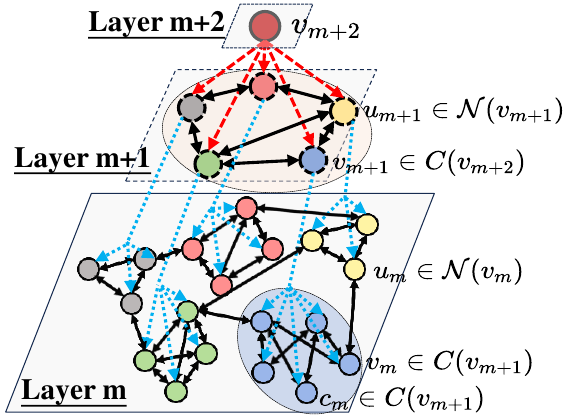}
	\caption{The hierarchical messaging mechanism.}
	\label{lu6}
\end{figure}

The input data of each node (child and parent)  is patched to obtain the initial feature embedding. Next, we use the extracted features as the representations of nodes in our multi-layer graph and conduct the message passing in GNN. Previous studies use GNNs to model the correlations between nodes at the same hierarchical level, neglecting potential hierarchical information. In this paper, we introduce the hierarchical message-passing to model potential hierarchical relationships among multiple passenger flow sequences. As illustrated in Fig.~\ref{lu6}, for a node $v_m$, the parent nodes are denoted as ${v_{m + 1}}$ and ${v_{m + 2}}$, the connected nodes is ${u_m}$, and the node ${c_m}$ has the same parent node as $v_m$. Firstly, the horizontal message passing is conducted within each layer (e.g. between $v_{m+1}$ and $u_{m+1}$ in the $(m+1)^{th}$ layer and $u_m$ and $c_m$ in the $m^{th}$ layer.). Subsequently, the hierarchical message passing is performed across layers (e.g. from $v_{m+2}$ to $u_{m+1}$ and $v_{m+1}$ in the $(m+1)^{th}$ layer, and from $u_{m+1}$ to $u_m$ and $v_{m+1}$ to $v_m$ and $c_m$ in the $m^{th}$ layer.).

\noindent \textbf{Horizontal Message Passing:} After extracting the local and global features from the input passenger flow sequence, we first perform the horizontal message passing in GNN. For a node $v_m$, we aggregate the information from its connected nodes using an aggregation function $\rho$. The aggregated information is then combined with the current node's features, and an update function $\phi$ is used to generate the new feature representations:
\begin{subequations}
\begin{align}
&{\bf{h}}({v_m}) = \phi \left( {{x^p}({v_m}),\rho \left( {{x^p}({u_m}),\left\{ {{\rm{\forall }}{u_m} \in {\cal N}({v_m})} \right\}} \right)} \right) \in {{\bf{R}}^{1 \times A * D}} \\
&{\bf{h}}({v_{m + 1}}) = \phi \left( {{x^p}({v_{m + 1}}),\rho \left( {\left\{ {{x^p}({u_{m+1}}),{\rm{\forall }}{u_{m+1}} \in {\cal N}({v_{m+1}})} \right\}} \right)} \right) \in {{\bf{R}}^{1 \times A * D}} \\
&{\bf{h}}({v_{m + 2}}) = \phi \left( {{x^p}({v_{m + 2}})} \right) \in {{\bf{R}}^{1 \times A * D}} 
\end{align}
\label{eq6}
\end{subequations}
\noindent \textbf{Hierarchical Message Passing:} After updating the nodes representation in each layer, we propagate the information of the parent nodes to the child nodes. To update the state of node $v_m$ through the hierarchical graph structure, we use the node ${v_{m + 2}}$ to update the state of node ${v_{m + 1}}$. Then, the updated state of node ${v_{m + 1}}$ is used to update the state of node $v_m$. Specifically, a learnable weight matrix is employed to allocate the features from parent nodes to child nodes, which can measure the hierarchical relationship between the parent node and different child nodes. Then, a function $\phi$ is applied for node state updating. The process is detailed as follows:
\begin{subequations}
\begin{align}
&{\bf{h'}}({v_{m + 1}}) = \phi \left( {{\bf{h}}({v_{m + 1}}),{M^{m + 2 \to m + 1}}{\bf{h}}({v_{m + 2}})} \right)\\
&{\bf{h'}}({v_m}) = \phi \left( {{\bf{h}}({v_m}),{M^{m + 1 \to m}}{\bf{h'}}({v_{m + 1}})} \right) 
\end{align}
\label{eq7}
\end{subequations}

Here, $M^{m + 2 \to m + 1}$ represents the weight matrix for representation from node ${v_{m + 2}}$ to node ${v_{m + 1}}$.
\subsection{Hierarchical Coordination}
After extracting the hierarchical relationships between different level nodes using GNN, a Gated Recurrent Unit (GRU) model is employed to extract the temporal features as its lightweight and comparable performance to LSTM. Additionally, when the task involves HP, further coordination is needed, as detailed below.

\noindent \textbf{Temporal feature extraction:} Inspiration from previous research, passenger flow sequences of nodes within the same cluster have the same model parameters. Therefore, for a node $v_m$ and its cluster nodes $c_m$, a GRU model is used to capture the temporal trends:
\begin{subequations}
\begin{align}
& g({v_m}) = GRU({\bf{h'}}({v_m})) \in {{\bf{R}}^{1 \times T}} \\
& g({c_m}) = GRU({\bf{h'}}({c_m}))
\end{align}
\label{eq8}
\end{subequations}

The initial predicted passenger flow for node $v_m$ is denoted as ${\tilde x^{t + 1:t + L}}({v_m}) = g({v_m})$. Without loss of generality, the initial predicted passenger flow matrix is denoted as ${\hat X^{t + 1:t + L}}({V_{Pr}}) = \bigcup\limits_v {g(v)} ,\forall v \in {V(Pr)}$, ${\hat X^{t + 1:t + L}}({V_{Pr}}) \in {{\bf{R}}^{V_{Pr} \times T}}$, where $V_{Pr}$ is the node set in prediction layer. When it is a TP task, we output the ${\hat X^{t + 1:t + L}}({V_{Pr}})$ as the final predicted passenger flow matrix.

\noindent \textbf{Coordination of Prediction Results:} In the case of the HP task, it is necessary to adjust the initial predicted passenger flows for all nodes in prediction layer to satisfy the hierarchical constraints. In coordination methods, optimization techniques are often employed to obtain a coordination matrix $P$. However, this approach has two drawbacks: (1) the obtained coordination matrix is not dynamic, and (2) it is challenging to integrate it into the DL model. Departing from the existing approach, we propose herein a learnable linear layer to approximate the coordination matrix $P$ \citep{burba2021trainable}, as illustrated in Fig.~\ref{lu7}. Specifically, we use the deep learning model MLP (Multilayer
Perceptron) to map all initial predicted values to the bottom node. The hierarchical matrix $H^r$ is then used to obtain the adjusted predicted values. Next, the observed values are compared to obtain the loss value, which is called the "Forward Propagation of Signals" process. The parameters in the MLP model are adjusted through the loss value, which is similar to the process of "Back Propagation of Errors". In the testing set, we will use the trained MLP model as a coordination matrix $P$. The detailed process is as follows:
\begin{figure}[ht]
	\centering
	\includegraphics[width=12cm]{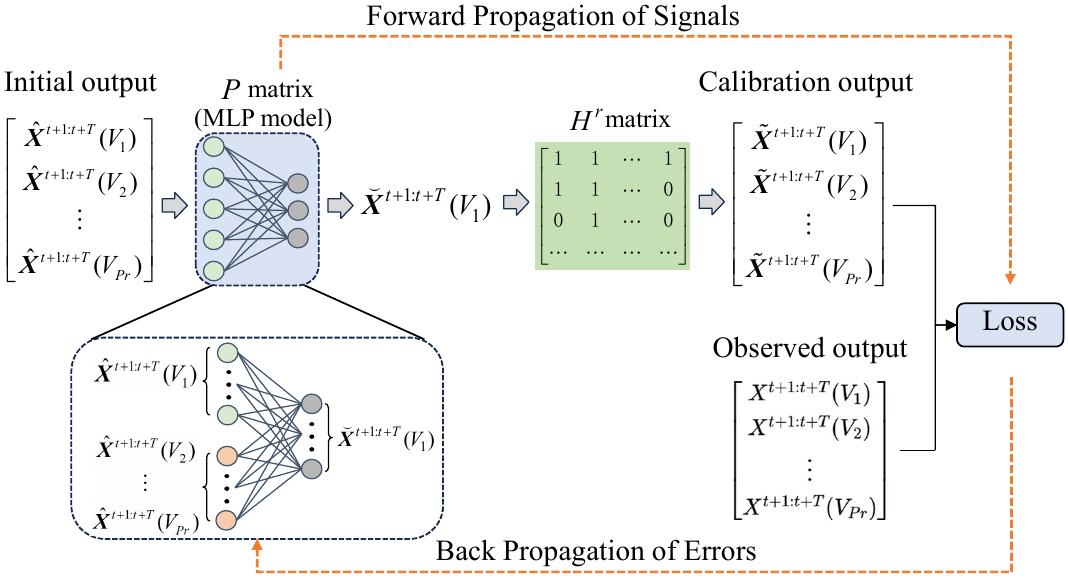}
	\caption{Hierarchical coordination process.}
	\label{lu7}
\end{figure}

First, by applying a linear layer, the initial predicted passenger flows of all prediction layers are transformed to the bottom layer. Then, the hierarchy matrix $H^r \in {{\bf{R}}^{V_{Pr} \times V_1}}$ is used to obtain the adjusted predicted passenger flows:
\begin{subequations}
\begin{align}
&{\mathord{\buildrel{\lower3pt\hbox{$\scriptscriptstyle\smile$}} 
\over X} ^{t + 1:t + T}}({V_1}) = MLP^{{\bf{R}}^{V_{Pr} \times T}\rightarrow {\bf{R}}^{V_1 \times T}}({\hat X^{t + 1:t + T}}({V_{Pr}})) \\
&{\tilde X^{t + 1:t + T}}({V_{Pr}}) = H^r \times {\mathord{\buildrel{\lower3pt\hbox{$\scriptscriptstyle\smile$}}  
\over X} ^{t + 1:t + T}}({V_1}) 
\end{align}
\label{eq9}
\end{subequations}

In order to mitigate the impact of outliers, the smooth quadratic loss (SQL) function \citep{mo2023timesql} is employed to calculate the error. The loss for node $v$ is given by:
\begin{align}
los{s_v} = \frac{1}{T}\sum\limits_{i = 1}^{T- 1} {\frac{{{{({{\tilde x}^{t + i:t + i + 1}}(v) - {x^{t + i:t + i + 1}}(v))}^2}}}{{{{({{\tilde x}^{t + i:t + i + 1}}(v) - {x^{t + i:t + i + 1}}(v))}^2} + z}}} 
\label{eq10}
\end{align}

where $z$ is a hyperparameter, taking values between 0 and 1. Finally, the loss is:
\begin{align}
loss = \frac{1}{{V_{Pr}}}\sum\limits_{v = 1}^{V_{Pr}} {los{s_v}} 
\label{eq11}
\end{align}
\subsection{Model learning}
\begin{figure}[ht]
	\centering
	\includegraphics[width=11cm]{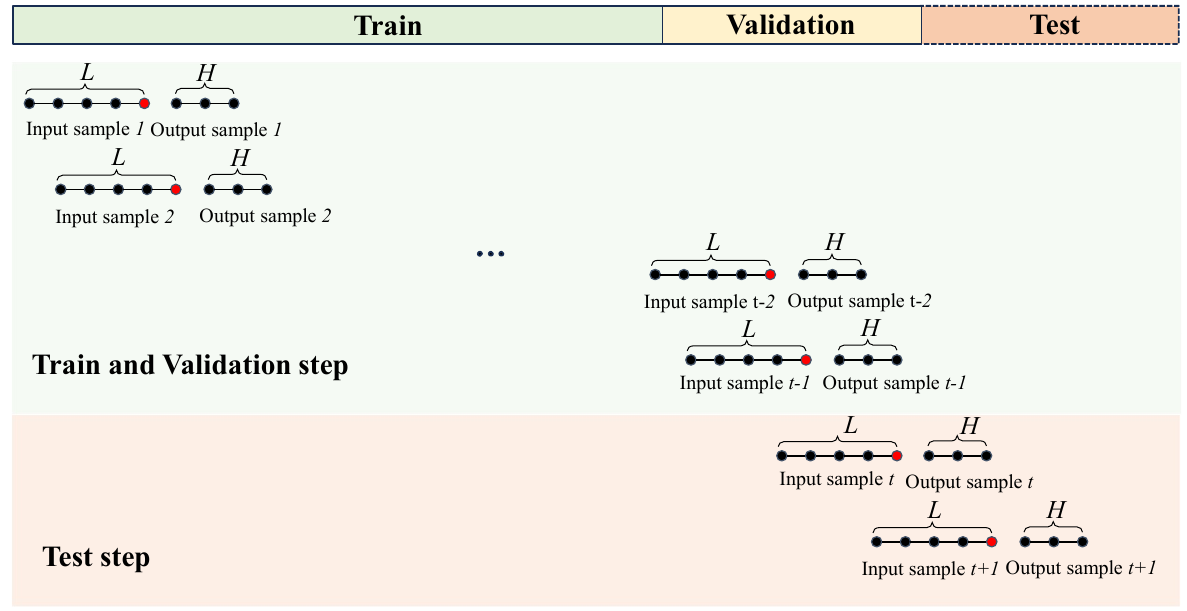}
	\caption{Data construction during training and testing phases.}
	\label{lu8}
\end{figure}

\noindent \textbf{(1) Sample Creation:} In this paper, we use the sliding window scheme to create the input and output passenger flow data, which serves as an effective data augmentation mechanism. Figure~\ref{lu8} illustrates the details of this scheme, where each iteration takes an input sample and generates an output sample. In the next time step, the sliding window moves forward by one time slots to create the new input and output samples.

\begin{figure}[ht]
	\centering
	\includegraphics[width=14cm]{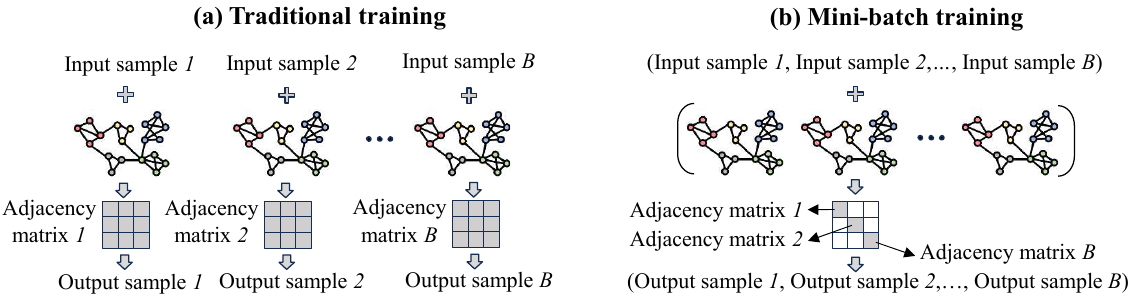}
	\caption{Mini-batch training of IPF-HMGNN.}
	\label{lu9}
\end{figure}

\noindent \textbf{(2) Model Training and Inference:} In our proposed IPF-HMGNN framework, a mini-batch strategy has been designed to accelerate the model training and inference. As depicted in Fig.~\ref{lu9}, multiple samples' graphs are combined into a large graph with no connections within a batch. The diagonal elements of the adjacency matrix represent the adjacency matrices of individual samples. Correspondingly, the inputs and outputs of each sample are concatenated, enabling the model to train on multiple samples simultaneously. Finally, the loss for all samples in a batch is averaged. Algorithm~\ref{algorithm1} provides the detailed steps for training of IPF-HMGNN.
\begin{algorithm}[ht]
	\caption{The process of training IPF-HMGNN}\label{algorithm1}
	\LinesNumbered 
	\KwIn{Different level graph $G$, history passenger flow matrix ${X^{t - L:t}}$, Random initial parameters $\theta$}
	\KwOut{Patching model parameters ${\theta ^p}$, GNN model parameters ${\theta ^g}$, GRU model parameters ${\theta ^r}$ and HP model parameters ${\theta ^h}$}
	\While{epoch $\leq$ Epochs;}
        {\For  {t $\in$ T}
        {Input $G_1$, $G_2$, $G_3$, ${X^{t - L:t}}$;\\
        Input series patching in all nodes and get the ${x^p}(v)$ by Eq~(\ref{eq4}),~(\ref{eq5});\\
        Horizontal Message Passing in all nodes and get the ${\mathop{\rm h}\nolimits} (v)$ by Eq~(\ref{eq6});\\
        Hierarchical Message Passing from top to bottom nodes and get the ${\mathop{\rm h}\nolimits} '(v)$ by Eq~(\ref{eq7});\\
        Temporal feature extraction and get the initial prediction by Eq~(\ref{eq8});\\
        Calculate the loss value by Eq~(\ref{eq10}) and ~(\ref{eq11});\\
        ${\theta ^p}$, ${\theta ^g}$, ${\theta ^r}$ $\gets$ $loss$;
        }}
    \eIf{Traditional task=True}{
     \Return{${\theta ^p}$, ${\theta ^g}$, ${\theta ^r}$} 
    }
    { 
     \While{epoch $\leq$ Epochs;}
        {\For  {t $\in$ T}
     {
     Input the  ${\hat X^{t + 1:t + L}}({V_{Pr}})$;\\
     Coordination of prediction results by Eq~(\ref{eq9});\\
     Calculate the loss value by Eq~(\ref{eq10}) and ~(\ref{eq11});\\
     ${\theta ^h}$  $\gets$ $loss$;
     }
     
     }
     \Return{${\theta ^p}$, ${\theta ^g}$, ${\theta ^r}$, ${\theta ^h}$}
    }
\end{algorithm}

\section{Experiment and Results}
\subsection{Experiment Settings}
\noindent \textbf{(1) Data Description:} We test the IPF-HMGNN in the Wixu metro system which consists of four lines and 44 stations. We use the outflow passenger data from March 1, 2017, to April 30, 2017, operating from 6:00 to 24:00 with a time granularity of 15 minutes. The dataset is divided into a training set (March 1 - April 20), a validation set (April 21 - April 25), and a test set (April 26 - April 30).

\noindent \textbf{(2) Programming Details:} We use the Deep Graph Library and the PyTorch deep learning framework for model construction. The hardware specifications include an R7-5800H CPU with a GTX 1650 GPU and 4GB memory. The specific parameters of IPF-HMGNN are performed using Optuna and as shown in Table~\ref{parameter}. 
\begin{table}[ht]
  \centering
  \caption{The parameter settings of the IPF-HMGNN.}
    \begin{tabular}{cccccc}
    \toprule
    Window size&Step size& Patch number & Depth convolution & Point convolution & Feature embedding\\
    \midrule
    $W$=36&$S$=8&$N$=8&$Q$=8&$A$=8&2*$T$ = 72 \\
    \midrule
    Batchsize& Dropout&Epochs&Learning rate& Look-back length & Prediction length\\
    \midrule
    64&0&100&0.001&$L$=72&$T$=36\\
    \bottomrule
    \end{tabular}
  \label{parameter}
\end{table}
\subsection{Model Evaluation and Comparison}
This study involves the regression task, and the model's accuracy is evaluated using Mean Absolute Error (MAE) and Root Mean Square Error (RMSE). Specifically:
\begin{subequations}
\begin{align}
&MAE = \frac{1}{{V_{Pr}}}\sum\limits_{v = 1}^{V_{Pr}} {MA{E_v}} ,MA{E_v} = \frac{1}{{T'}}\sum\limits_{t = 1}^{T'} {\left| {{x^t}(v) - {{\tilde x}^t}(v)} \right|} \\
&RMSE = \frac{1}{{V_{Pr}}}\sum\limits_{v = 1}^{V_{Pr}} {RMS{E_v}} ,RMS{E_v} = \sqrt {\frac{1}{{T'}}\sum\limits_{t = 1}^{T'} {{{({x^t}(v) - {{\tilde x}^t}(v))}^2}} }
\end{align}
\end{subequations} 

where ${x^t}(v)$ and ${\tilde x^t}(v)$ represent the actual and predicted values, respectively; $\bar x(v)$ is the mean value; $T'$ is the sample size of the test set.

In addition, for hierarchical prediction task, we create a hierarchical error metric between node ${v_{m + 1}}$  and its child nodes ${v_m} \in C({v_{m + 1}})$:                                                                    
\begin{align}
{E^{m + 1}}(t) = \left( {\sum\limits_{{v_m} \in C({v_{m + 1}})} {{{\tilde x}^t}({v_m})} } \right) - {\tilde x^t}({v_{m + 1}})
\end{align}

\begin{figure}[ht]
	\centering
	\includegraphics[width=10cm]{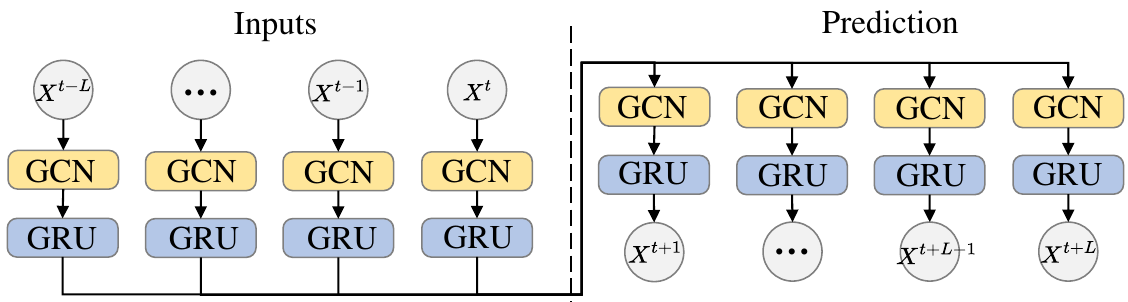}
	\caption{The structures of the T-GCN model.}
	\label{lu10}
\end{figure}

A baseline model system is constructed to evaluate the model's utility, incorporating the TP and HP approaches. The TP approaches include historical average (HA) and commonly used DL models, such as the MLP (Multilayer Perceptron), LSTM, GRU, and other time series prediction models. Additionally, the temporal graph convolutional network (T-GCN) \citep{8809901} is used for spatiotemporal modeling, as shown in Fig.~\ref{lu10}. We employ the GCN and Graph Attention Network (GAT) as the GNN model in T-GCN, and denoted as T-GAT. Then, the GCN, GAT, Spectral Graph Convolution (SGC), Graph Isomorphism Network (GIN), and Sample and Aggregation (SAGE) are implemented in the IPF-HMGNN. The HP approach includes bottom-up (BU), middle-out (MO), and top-down (TD) methods. For detailed information about these models, please refer to Appendix \textbf{\hyperlink{Appendix A}{A}} and \textbf{\hyperlink{Appendix B}{B}}.

\subsection{Analysis of Prediction Results} \label{5.3}
We conduct extensive experiments to obtain insights into the followings: \textbf{\hypertarget{(1)}(1)} impact of the different parameters on the performance of the proposed IPF-HMGNN; \textbf{\hypertarget{(2)}(2)} impact of hierarchy selection on the proposed IPF-HMGNN performance; \textbf{\hypertarget{(3)}(3)} effectiveness of the different GNN models; \textbf{\hypertarget{(4)}(4)} influence of the HP coordination module on the prediction errors; \textbf{\hypertarget{(5)}(5)} effectiveness of the proposed patch operation to extract local and global features in historical passenger flows; \textbf{\hypertarget{(6)}(6)} effectiveness of the hierarchical message-passing mechanism in capturing the hierarchical relationships between different layer graphs; \textbf{\hypertarget{(7)}(7)} application and effectiveness of the proposed IPF-HMGNN in existing passenger flow prediction models.

To obtain the above insights, two applications are created in our numerical analysis below. The Application 1 involves passenger flow TP tasks without the requirement of meeting the hierarchical constraints. The Application 2 is the passenger flow HP task, where predictions for various passenger flow sequences must satisfy the hierarchical constraints. 
\subsubsection{Application 1: Traditional Prediction}
For the TP task, we conduct two prediction scenarios: (1) the station passenger flow prediction and (2) the station and per ticket type passenger flow prediction. 

\begin{figure}[ht]
	\centering
	\includegraphics[width=15cm]{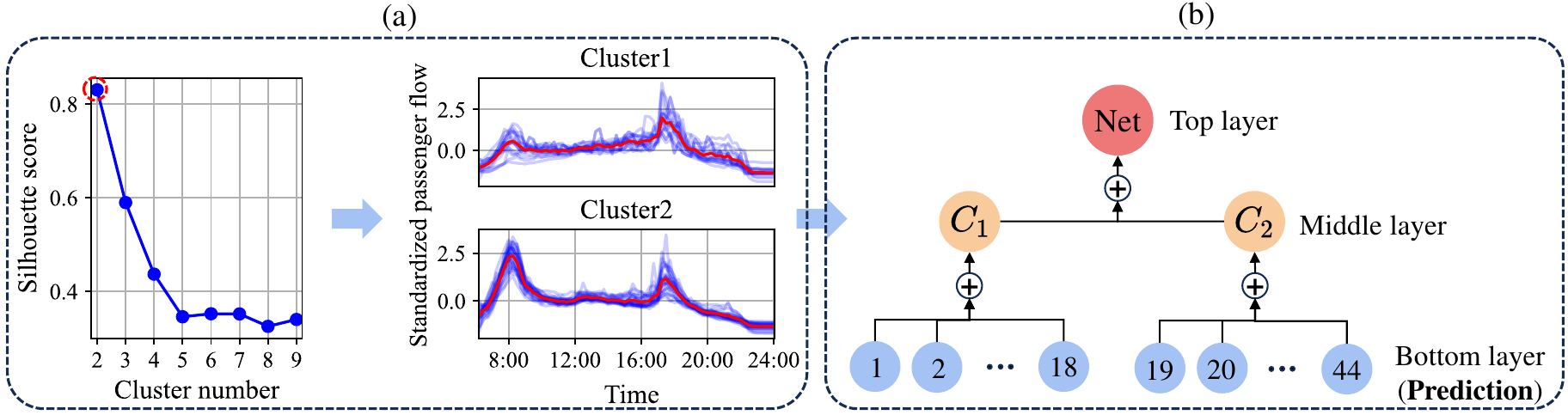}
	\caption{Hierarchical structure construction process for station passenger flow prediction, (a) clustering process; (b) hierarchical construction.}
	\label{lu11}
\end{figure}

To deal with the complexity and diversity of passenger flows, researchers often cluster the stations based on passenger flow data, where the stations in same cluster have the similar passenger flow characteristics with the same model parameters. For TP task, we employ the popular $k-means$ clustering algorithm \citep{FANG2024122550} to construct a hierarchical structure, as explained in Sec \ref{4.2} and illustrated with the actual data collected from the Wixu metro system in Fig.~\ref{lu11}. The cluster 1 exhibits the significant morning and evening peak periods, with the evening peak outflow exceeding the morning peak outflow. This cluster represents the residential and entertainment-oriented stations. The cluster 2 also shows significant morning and evening peak periods, but the morning peak outflow is greater than the evening peak outflow, representing the employment-oriented stations. In Fig.~\ref{lu11}.b, the bottom layer contains the individual stations, the middle layer represents the station clusters and the top layer represents one node with aggregated flows from stations. The prediction target is the passenger flow of individual stations in the bottom layer.

\begin{figure}[ht]
	\centering
	\includegraphics[width=15cm]{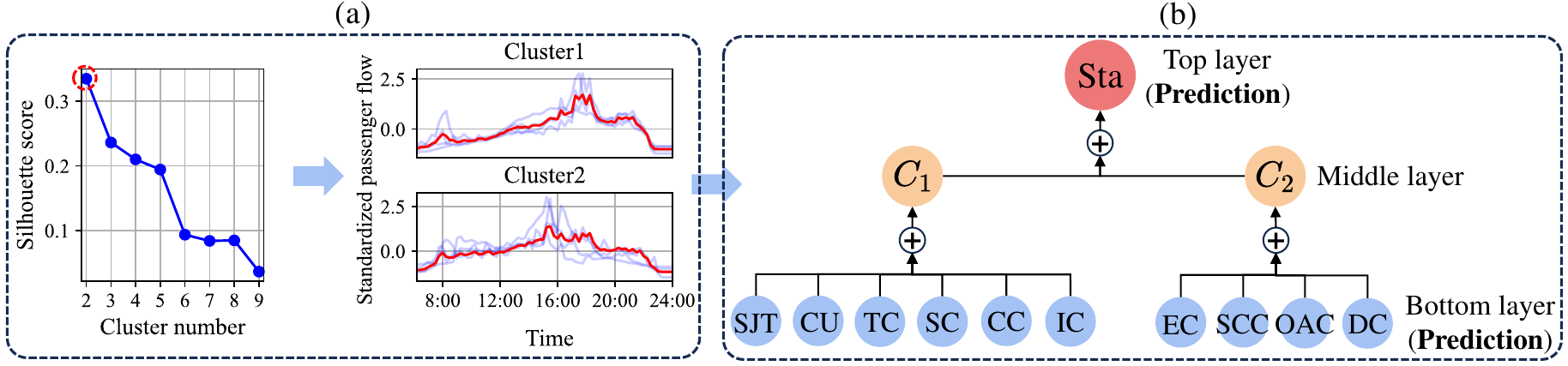}
	\caption{Hierarchical structure construction process for station and per ticket type passenger flow prediction, (a) clustering process; (b) hierarchical construction; SJT: one-way ticket; CU: city joint card; TC: time card; SC: student card; CC: citizen Card; IC: financial IC card; EC: employee card; SCC: free senior citizen card; OAC: elderly card; DC: disabled card.}
	\label{lu12}
\end{figure}

The proposed IPF-HMGNN can also be used to predict the single-station (Interchange station) passenger flow and the passenger flow per ticket type at the station, as shown in Fig.~\ref{lu12}. We first group via clustering the passenger flow data sequences per ticket types. The cluster 1 includes one-way tickets, student cards, citizen cards, etc., concentrating in one prominent peak periods. The cluster 2 includes elderly free cards, senior citizen cards, disability cards, etc., with passenger flows having a more extended peak period. Figure~\ref{lu12}.b depicts the constructed hierarchical structure. The bottom layer contains the passenger flows per ticket types. The middle layer represents the cluster passenger flows and and the top layer represents the aggregated station passenger flows.

The above models are then used to make predictions in the two scenarios described previously, and the results are shown in Tables~\ref{tab2} and~\ref{tab3}. The optimal metrics are highlighted in bold black, while the second-best metrics are underlined. It can be observed that the prediction errors of GNN models such as T-GCN and T-GAT are smaller than those of traditional time series prediction models like MLP and LSTM. This is mainly attributed to the ability of the GNN models to capture the correlations between passenger flow sequences. Furthermore, the GNN models trained under the IPF-HMGNNsignificantly reduce the prediction errors. The results indicate the effectiveness of the proposed IPF-HMGNN.
\begin{table}[ht]
  \centering
  \caption{Prediction errors of different methods in station passenger flow prediction scenarios, +IPF means training under the IPF-HMGNN framework.}
    \begin{tabular}{ccccccc}
   \toprule
   Target& \multicolumn{2}{c}{Ordinary station} & \multicolumn{2}{c}{Interchange station} &\multicolumn{2}{c}{All station} \\ 
   \midrule
   Method&MAE	&RMSE	&MAE&	RMSE	&MAE	&RMSE\\
    \midrule
    HA	&16.103&	23.977	&127.347&	197.625&19.807&	29.739\\
   MLP&	14.172&	21.069	&103.155&	154.871	&16.660&	24.888\\
   LSTM	&13.120 &	20.469& 	97.274 &	139.828 &	15.425 &	23.140 \\
   GRU	&13.126 &	20.223 &	99.731& 	153.439 	&15.527& 	23.091 \\
   T-GCN	&13.118	&19.498	&96.017	&141.972&	15.518&	23.096\\
   T-GAT&	12.971	&19.526&	76.294	&114.935	&14.938	&22.525\\
   GCN+IPF&	\textbf{11.465}	&\textbf{16.827}	&\textbf{48.429}&\textbf{65.480}&\textbf{13.456}&\textbf{19.712}\\
   GAT+IPF	&\underline{11.561}&\underline{17.147}	&\underline{52.485}&\underline{75.304}	&\underline{13.765}&\underline{20.482}\\
   \midrule
   Improvement (GCN) \%&	12.601&	13.699	&49.562	&53.878&	13.288&	14.652\\
   Improvement (GAT) \%	&10.870	&12.184&	31.207&	34.481&	7.852&	9.070\\
   \bottomrule
    \end{tabular}
     \begin{tablenotes}
        \footnotesize
        \item[1] Note: Improvement is the error reduction rate of GNN model trained under the IPF-HMGNN.
      \end{tablenotes}
  \label{tab2}
\end{table}

\begin{table}[ht]
  \centering
  \caption{Prediction errors of different methods for per ticket types passenger flow prediction scenarios, +IPF means training under the IPF-HMGNN framework.}
    \begin{tabular}{ccccccccc}
   \toprule
   Target& \multicolumn{2}{c}{SJT} & \multicolumn{2}{c}{CC} &\multicolumn{2}{c}{All card} &\multicolumn{2}{c}{Interchange station}\\ 
   \midrule
   Method&MAE	&RMSE	&MAE&	RMSE	&MAE	&RMSE&MAE	&RMSE\\
    \midrule
    HA&	81.664	&129.187&	47.714	&70.573	&15.380	&23.508	&127.347	&197.625\\
MLP	&46.846&	79.902&	31.679&	48.062	&10.192&	16.251	&76.839	&125.104\\
LSTM&	43.243	&75.057	&29.498	&46.418&	9.461	&16.430	&72.286	&102.592\\
GRU	&42.864	&77.161	&29.844&	47.734 &	9.644	&15.131	&69.160&	101.052\\
T-GCN	&42.039	&71.721	&32.753	&48.222&	9.561	&15.008&	57.479	&91.136\\
T-GAT	&42.868	&74.767&	27.763	&40.029&	9.145&	14.506	&60.015	&89.244\\
GCN+IPF&\textbf{28.927}	&\underline{49.576}	&\underline{25.928}	&\underline{35.938}	&\textbf{7.569}&\underline{11.562}	&\textbf{46.400}&\underline{68.429}\\
GAT+IPF	&\underline{31.018}&\textbf{48.318}&\textbf{23.986}&\textbf{33.409}&\underline{7.582}&	\textbf{11.151}&\underline{46.823}&\textbf{66.869}\\
\midrule
Improvement (GCN) \%&	31.190	&30.877	&20.838	&25.474	&20.835	&22.961	&31.190	&30.877\\
Improvement (GAT) \%&	27.643&	35.375	&13.604	&16.538&	17.091	&23.128&	27.643&	35.375\\
   \bottomrule
    \end{tabular}
  \label{tab3}
\end{table}

\begin{figure}[ht]
	\centering
	\subfigure[]{
		\includegraphics[width=11.0cm]{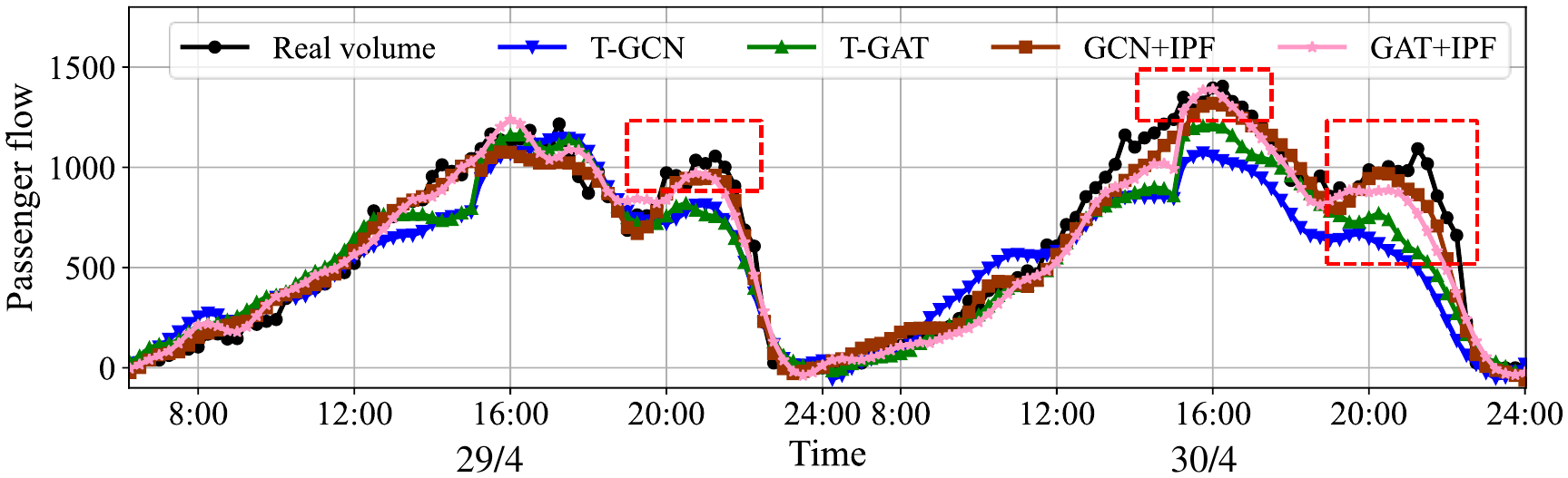}
	}
	\subfigure[]{
		\includegraphics[width=11.0cm]{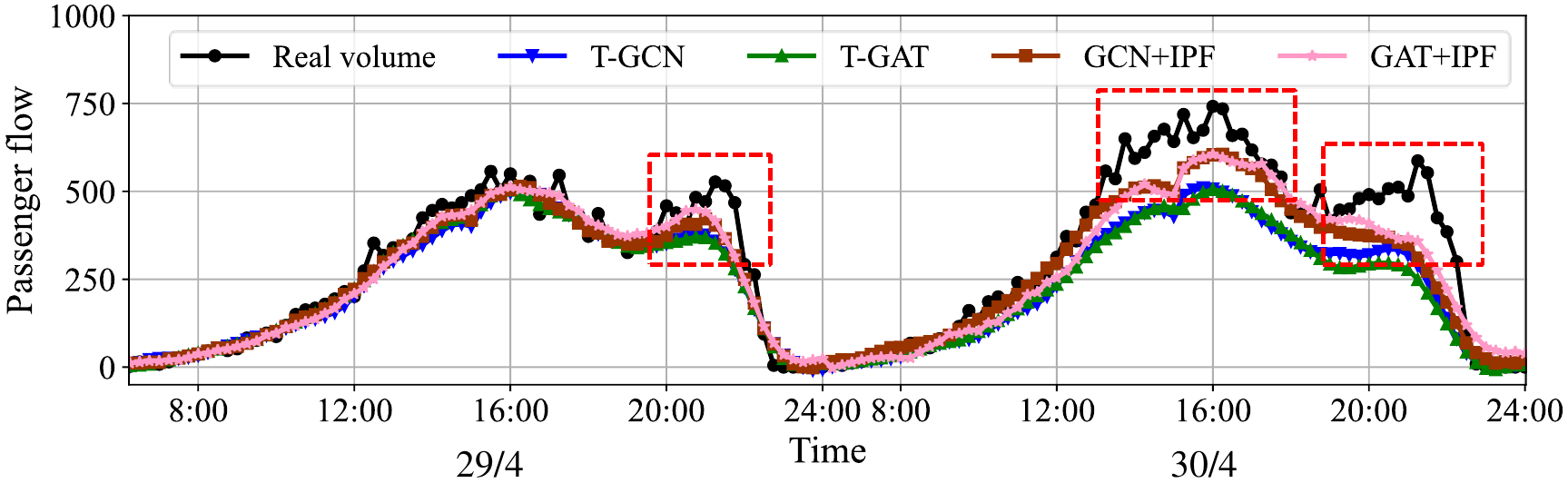}
	}
	\caption{Visualization of the GNN model prediction results, (a) is the prediction passenger flow of Interchange station, and (b) is the prediction passenger flow of SJT card.}
	\label{lu13}
\end{figure}

Furthermore, we visualize the prediction results of the GNN models in Fig.~\ref{lu13}. It can be seen that the prediction values of the GNN models trained under the IPF-HMGNN are much closer to the actual values, especially during the off-peak hours, which indicates the effective modeling ability of IPF-HMGNN for off-peak hours.

\begin{figure}[ht]
	\centering
	\includegraphics[width=14cm]{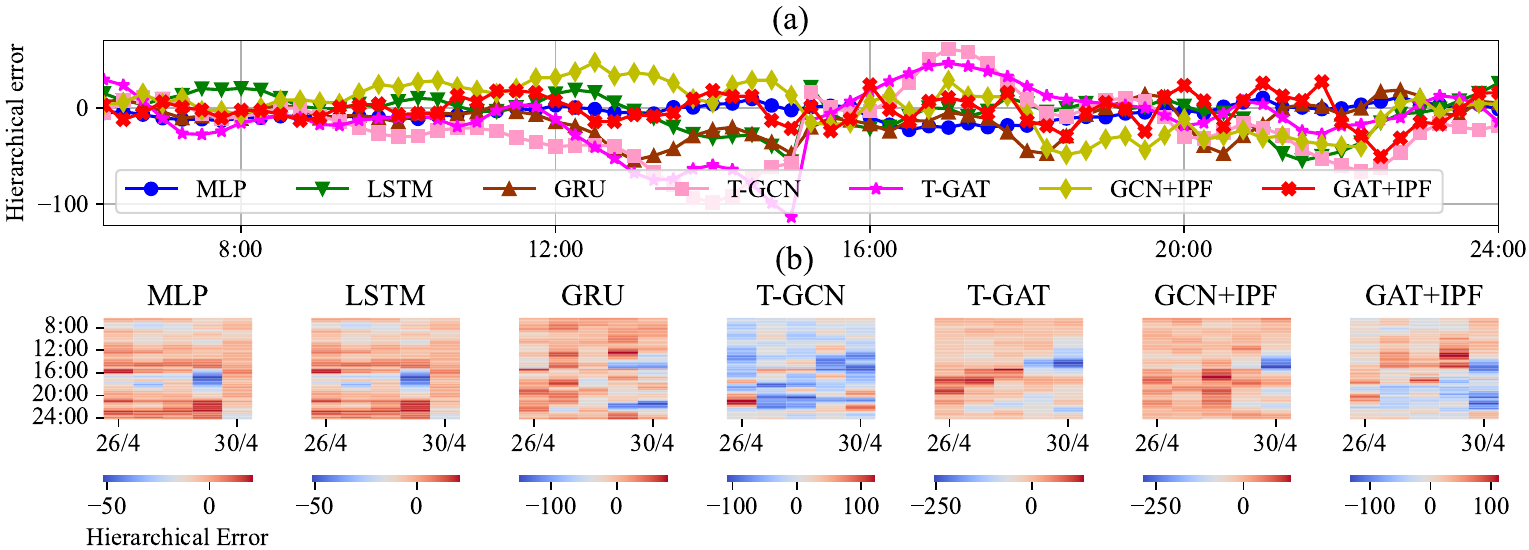}
	\caption{The hierarchical error at different times, (a) is the average hierarchical error of the test set, and (b) is the hierarchical error on different dates.}
	\label{lu14}
\end{figure}

When predicting passenger flow per ticket types in TP task, it is necessary to simultaneously predict both the passenger flows for the station and that per ticket type. The sum of the predicted values per ticket type must also be equal to the station passenger flow. However, it can be observed in Fig.~\ref{lu14} that the sum of predicted passenger flow per ticket types differs significantly from the predicted value of the station. The hierarchical errors occur during off-peak hours, and the better the performance of the models is, the larger hierarchical errors are. This will be tackled in the HP task below.
\subsubsection{Application 2: Hierarchical Prediction}
To address the aforementioned hierarchical error issue, the adjustments should be made after predicting the passenger flows for stations and that per ticket type to meet the hierarchical constraint. This constraint ensures that the sum of predicted passenger flows per ticket type equals the predicted passenger flow for the station. Results using our proposed IPF-HMGNN are shown in Table~\ref{tab4}. The following conclusions can be drawn:

1. The BU method produces smaller prediction errors when focusing more on the bottom-layer prediction targets. Hence, the BU method ensures that bottom-layer nodes are not influenced by top-layer nodes.

2. When emphasizing the top-layer prediction targets, the TD method is recommended. The TD involves univariate passenger flow prediction for top-layer nodes (one node), significantly improving model training efficiency. On the other hand, using BU to obtain the top-layer node prediction passenger flow often leads to the worse results.

3. The IPF-HMGNN can balance the  prediction errors for bottom- and top-layer prediction targets, achieving the good predictive performance at both layers.
\begin{table}[ht]
  \centering
  \caption{Prediction errors of different methods in HP scenarios.}
    \begin{tabular}{lllllllll}
   \toprule
   
   Target& \multicolumn{2}{c}{SJT} & \multicolumn{2}{c}{CC} &\multicolumn{2}{c}{All card} &\multicolumn{2}{c}{Interchange station}\\ 
   \midrule
   Method&MAE	&RMSE	&MAE&	RMSE	&MAE	&RMSE&MAE	&RMSE\\
    \midrule
    \multicolumn{9}{l}{\textbf{BU}} \\
    \midrule
    HA-BU&	81.664&	129.187	&47.714	&70.573&	15.380&	23.508&	127.347	&197.625\\
MLP-BU&	47.700&	82.693&	33.137&	51.020&	10.431&	16.828&	81.607&	133.031\\
LSTM-BU	&44.258 &	75.840 	&30.935 	&42.987 	&9.198 	&14.054 &	77.867& 	123.176 \\
GRU-BU&	44.773 	&71.268 &	29.575 &	41.128 &	9.382& 	13.975 &	71.971 &	105.878 \\
T-GCN-BU&	42.107	&69.298	&26.614&	37.545	&8.914	&13.642	&64.769	&98.090\\
T-GAT-BU	&41.904	&73.804&	28.669	&41.019	&9.138	&14.491	&70.017&	111.623\\
\midrule
\multicolumn{9}{l}{\textbf{MO}} \\
\midrule
HA-MO&	81.493	&130.583&	47.599	&69.250&	15.361	&23.560&	127.347&	197.625\\
MLP-MO&	54.570&	89.360&	33.993&	50.841&	11.227	&17.531	&71.921	&111.177\\
LSTM-MO&	50.382 &	86.099 	&33.606 	&49.637 &10.790 &16.991 &	67.788 &	96.677 \\
GRU-MO	&52.884 &	87.200 &	32.145 &	49.751 &	10.875 	&17.140 &	67.703 &	102.644 \\
T-GCN-MO&	50.338	&84.092	&30.712&	48.513	&10.375	&16.596&	62.429	&95.280\\
T-GAT-MO	&51.447&	91.212	&29.982	&46.013&	10.408	&17.055&	65.661	&111.386\\
\midrule
\multicolumn{9}{l}{\textbf{TD}} \\
\midrule
HA-TD&	81.400&	130.864&	47.564&	69.620&	15.383&	23.707	&127.350	&197.625\\
MLP-TD&	50.281	&84.265	&32.210&	48.673	&10.986	&18.064	&60.215&	92.873\\
LSTM-TD	&46.511 &	77.941 &	31.328 &	48.916 	&10.591 &	17.334 &	54.228 &	82.802\\ 
GRU-TD	&47.676 	&79.656 	&31.103 	&47.578 &10.663 &	17.316 	&56.342 &	85.598\\ 
T-GCN-TD&	46.864&	76.838	&30.330&	46.756&	10.360&	16.763	&54.240 &	79.045 \\
T-GAT-TD	&45.314&	74.307	&32.681	&50.621	&10.535	&17.164	&54.965 	&80.663 \\
\midrule
\multicolumn{9}{l}{\textbf{+IPF}} \\
\midrule
GCN+IPF	&\textbf{27.233}&\textbf{46.797}&\underline{23.521}&\underline{32.983}	&\textbf{7.150}&\textbf{10.975}&\textbf{46.745}	&\underline{73.160}\\
GAT+IPF&\underline{29.820}&\underline{47.104}&\textbf{21.611}&\textbf{30.643}&\underline{7.363}	&\underline{11.035}&\underline{47.585}&\textbf{68.148}\\
\midrule
Improvement (GCN) \%&	35.324	&32.470&	11.622	&12.151	&19.789&	19.550&	13.818	&7.445\\
Improvement (GAT) \%&	28.837	&36.177	&24.619	&25.296&	19.424	&23.849&	13.427&	15.515\\
   \bottomrule
    \end{tabular}
    \begin{tablenotes}
        \footnotesize
        \item[1] Note: Improvement is the error reduction rate of GNN trained under the IPF-HMGNN compared to the traditional optimal hierarchical model GNN.
      \end{tablenotes}
  \label{tab4}
\end{table}
\subsubsection{Performance Impact Analysis}
\begin{figure}[ht]
	\centering
	\includegraphics[width=12cm]{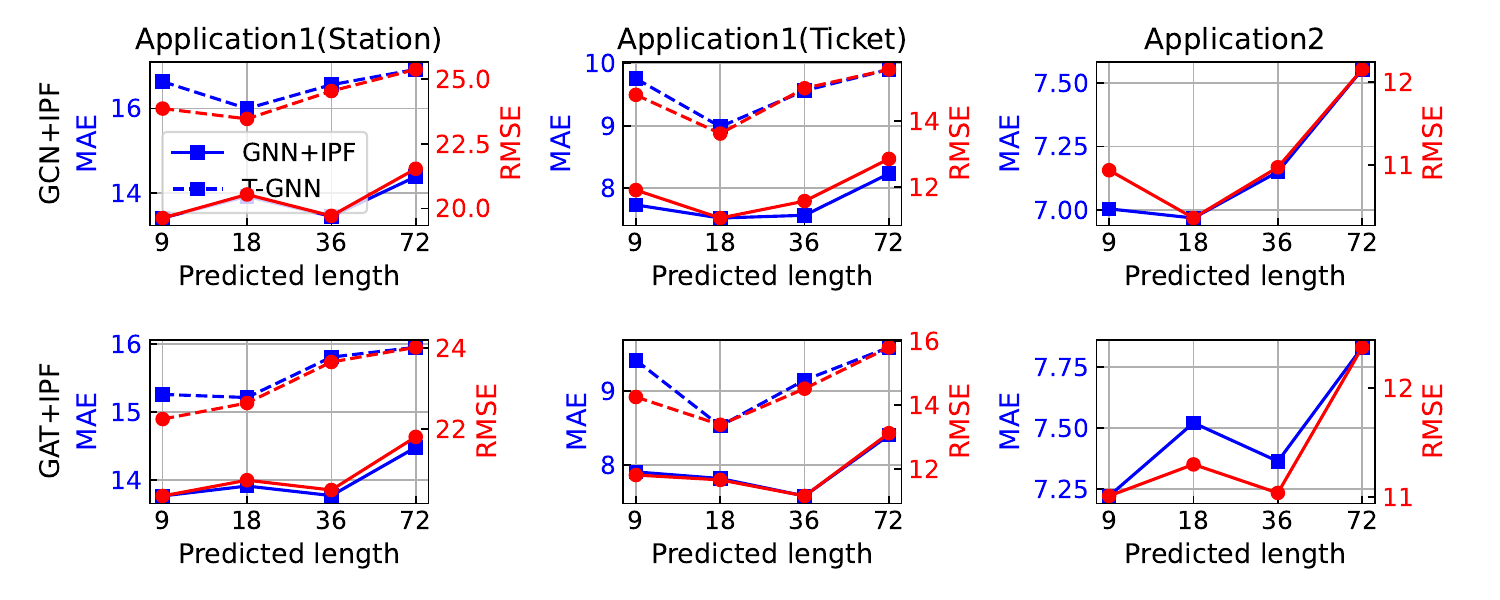}
	\caption{Prediction error of IPF-HMGNN at different prediction lengths.}
	\label{lu15}
\end{figure}

To study the \textbf{\hyperlink{(1)}{1}}st aspect listed in Sec \ref{5.3}, we analyze the impact of the hyperparameters on the performance of the proposed IPF-HMGNN. These parameters include the prediction length $T$, the number of separation convolution channels $A$, and the $K$ value in constructing the bottom layer graph. Figure~\ref{lu15} presents the average prediction indicators for all stations and all ticket types under different prediction lengths. It can be observed that when the prediction length $T$ is below 72, the changes of MAE and RMSE are not significant. However, the prediction error significantly increases when the prediction length $T$ is 72. This is because the prediction length $T$ is greater than or equal to the lookback window length $L$. The model relies on limited historical passenger flow information and lacks sufficient features to capture the longer-term trends.

\begin{figure}[ht]
	\centering
	\includegraphics[width=12cm]{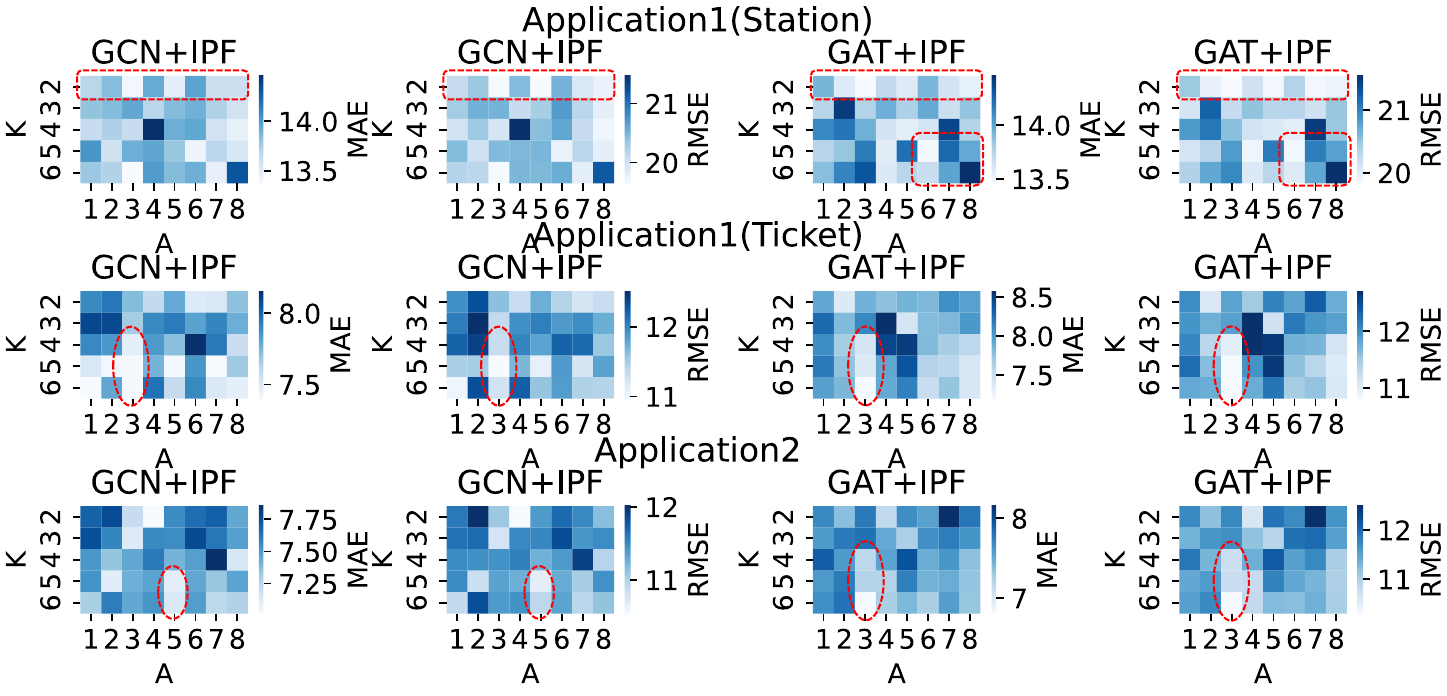}
	\caption{Prediction error of IPF-HMGNN at different $A$ and $K$ values.}
	\label{lu16}
\end{figure}

In addition, we analyze the impact of $A$ and $K$ in the IPF-HMGNN, as shown in Fig.~\ref{lu16}. When choosing different $A$ and $K$ values, the prediction errors of the IPF-HMGNN are consistently smaller than the baseline models. For station passenger flow prediction, smaller $K$ value is preferable. This is because station passenger flow is relatively stable. For per ticker type passenger flow prediction, larger $K$ value is preferable due to the high volatility of passenger flow per ticket type, making it challenging to predict. Increasing the number of connecting nodes allows the model to extract more feature information related to their passenger flow sequences.

\subsubsection{Hierarchy selection}
\begin{figure}[ht]
	\centering
	\includegraphics[width=12cm]{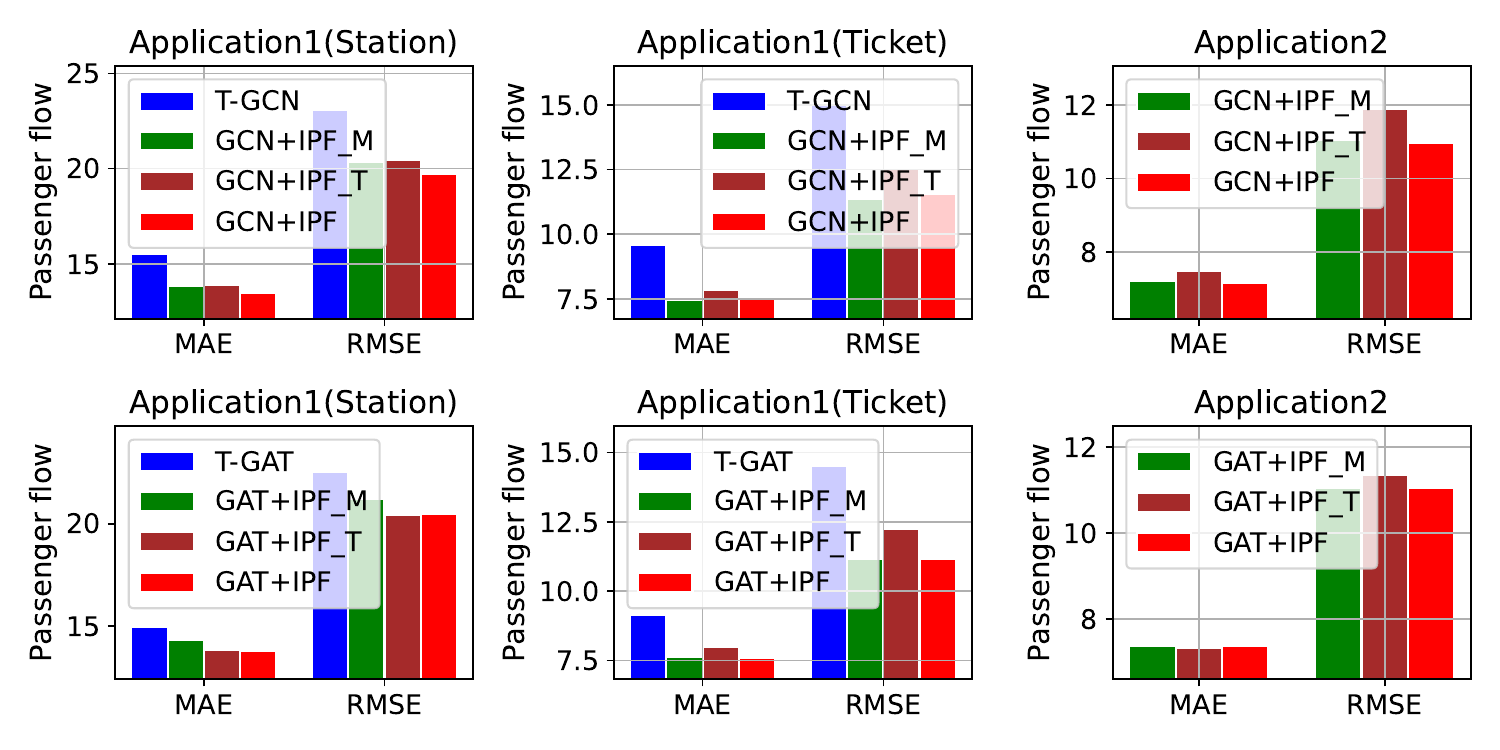}
	\caption{Prediction errors under different hierarchies; \text{+IPF\_T} and \text{+IPF\_M}  indicate using only top-layer or middle-layer nodes for message passing, respectively.}
	\label{lu17}
\end{figure}

For the \textbf{\hyperlink{(2)}{2}}nd aspect, we compare the prediction errors under different hierarchical structures. In the proposed IPF-HMGNN, recall that a hierarchical graph with three layers (bottom, middle, and top) is used. We use the middle-layer (\text{+IPF\_M}) or top-layer (\text{+IPF\_T}) nodes for information propagation and obtained the prediction errors as shown in Fig.~\ref{lu17}. It can be observed that using a portion or all hierarchical layers in the IPF-HMGNN reduces prediction errors. When using only middle-layer or top-layer nodes for message passing, the prediction errors of the IPF-HMGNN increase. This indicates that nodes at different hierarchical layers have different feature information, all contributing to the bottom-layer nodes' prediction task. Furthermore, \text{+IPF\_T} has more prediction errors than \text{+IPF\_M}, suggesting that middle-layer nodes are more crucial for the prediction task of the bottom-layer nodes. This is because the middle-layer nodes can better distinguish different passenger flow patterns than top-layer nodes. In conclusion, a multi-layer message-passing should be employed in practical applications, including nodes from different layers.
\subsubsection{Robustness Analysis} \label{5.3.5}
For the \textbf{\hyperlink{(3)}{3}}rd aspect, we analyze the prediction errors under different GNN models and bottom-layer graphs in the IPF-HMGNN.

\noindent \textbf{(1) GNN Adaptability Analysis}

\begin{figure}[ht]
	\centering
	\includegraphics[width=15cm]{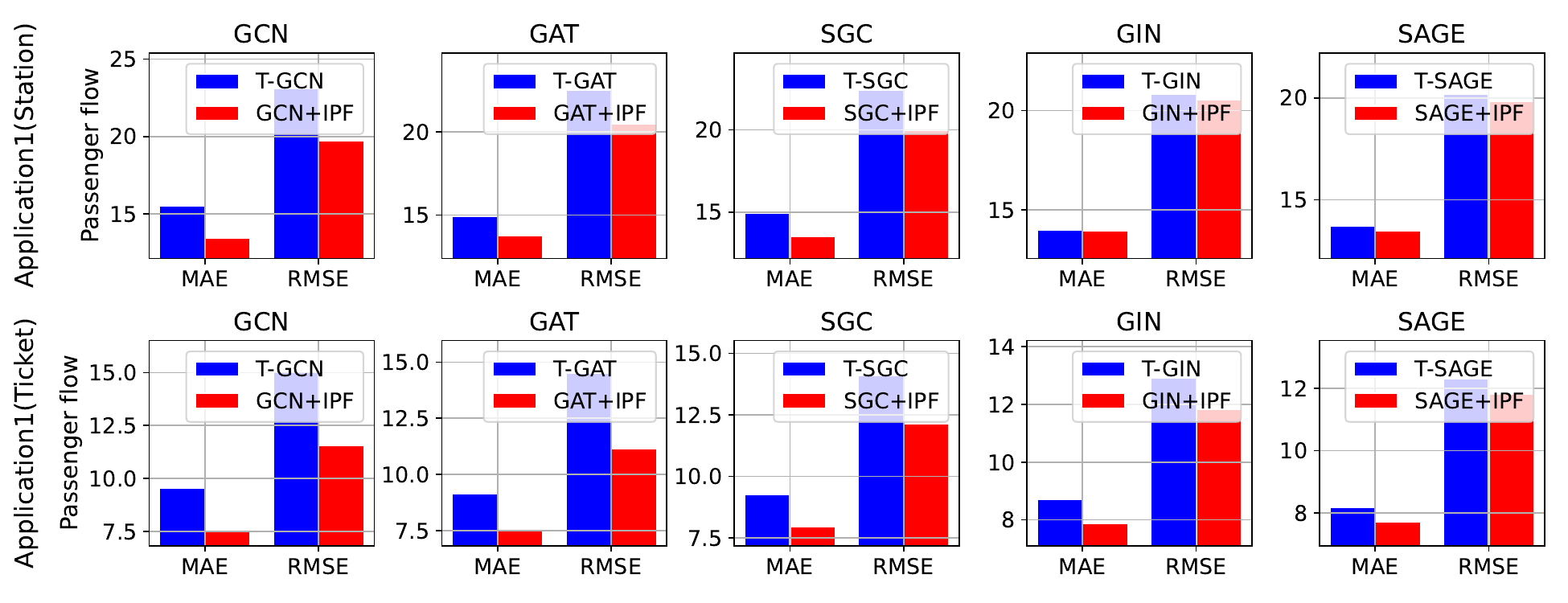}
	\caption{Prediction error of IPF-HMGNN under different GNN models.}
	\label{lu18}
\end{figure}

To validate the effectiveness of the IPF-HMGNN for different GNN models, the GCN, GAT, SGC, GIN, and SAGE models are applied. As shown in Fig.~\ref{lu18}, it can be observed that using different GNN models in the IPF-HMGNN significantly reduces errors. Moreover, compared to the station passenger flow prediction task, the benefits obtained for per ticket type passenger flow prediction are even more significant. This is because modeling the correlation between different ticket-type passenger flows is more complex and requires the feature information provided by the middle and top-layer nodes.

\noindent \textbf{(2) Hierarchical Robustness Analysis}

Furthermore, we execute prediction tasks using different bottom-layer graphs. For scenario 1, we construct the passenger flow correlation, clustering, physical topology, and random graphs. In scenario 2, there is no physical topology structure, we only use the passenger flow correlation, clustering, and random graphs. In the clustering graphs, nodes of the same cluster are connected, and the physical topology graph represents the actual connections in the metro network. In the random graph, the connections between nodes are obtained randomly using a uniform distribution method.

\begin{figure}[ht]
	\centering
	\includegraphics[width=13cm]{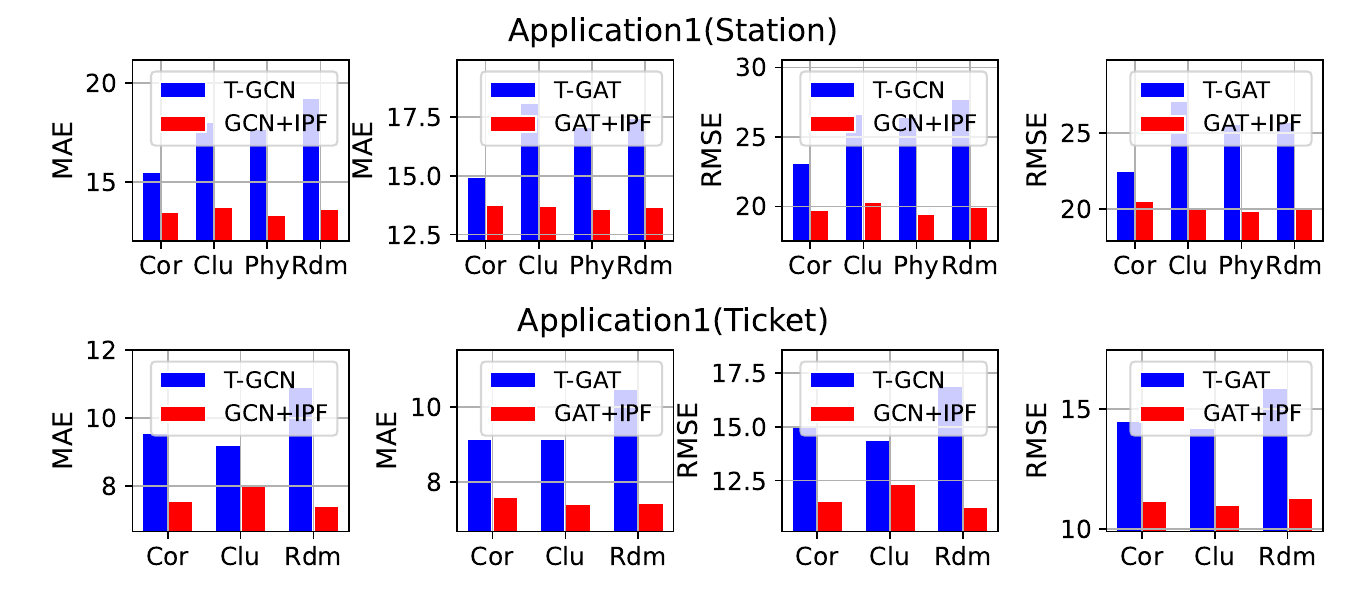}
	\caption{Prediction errors of different bottom layer graphs: Cor, Clu, Phy and Rdm represents the correlation graph, clustering graph, physical graph, and random graph, respectively.}
	\label{lu19}
\end{figure}

The results using different bottom-layer graphs are shown in Fig.~\ref{lu19}. It can be observed that when using different bottom graphs, the IPF-HMGNN can enhance the predictive performance of traditional GNN models. In addition, the predefined bottom graphs significantly impact the predictive performance of traditional GNNs. When applying a random graph in the GNN model for prediction, the performance is often subpar, indicating the effectiveness of predefined bottom-layer graphs. Moreover, the IPF-HMGNN shows slight variation in prediction errors under different bottom-layer graphs, demonstrating the stronger robustness. This is mainly because, in the IPF-HMGNN, bottom-layer nodes also integrate the information from the middle and top-layer nodes.
\subsubsection{HP Module Impact Analysis}
\begin{figure}[ht]
	\centering
	\includegraphics[width=14cm]{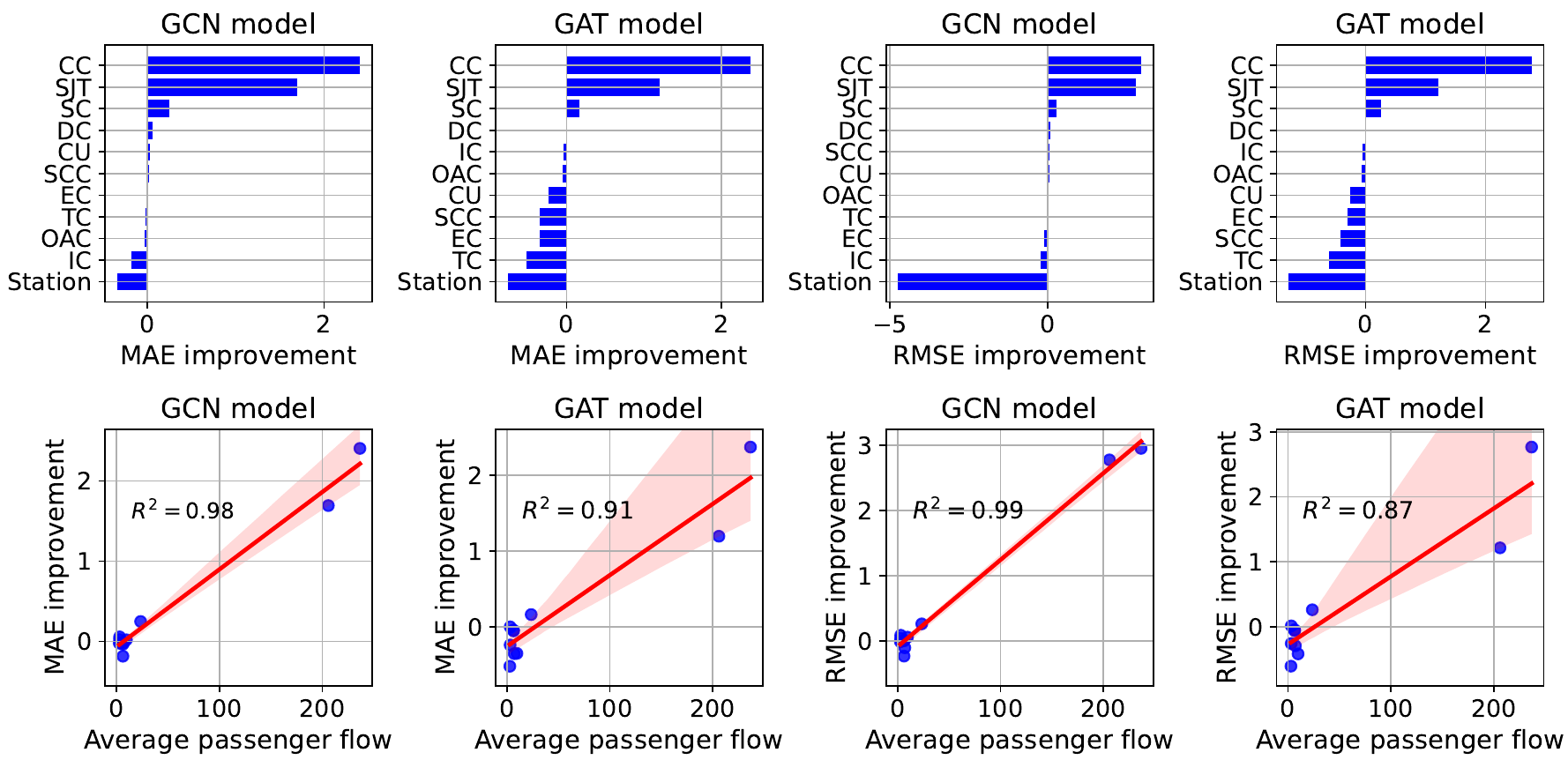}
	\caption{The impact of HP module on prediction errors of different passenger flow sequences.}
	\label{lu20}
\end{figure}

Addressing the insight \textbf{\hyperlink{(4)}{4}}th aspect, we investigate the influence of the added HP coordination module on the prediction errors. Firstly, we compute the difference in prediction errors between TP and HP tasks when simultaneously predicting both station and per ticket type passenger flows. Then, we calculate the averages of different passenger flow sequences in the test set. As shown in Fig.~\ref{lu20}, the x-axis of each subplot in the first row represents the change in prediction error for HP relative to TP approach. The x-axis of each subplot in the second row represents the average passenger flow for per ticket type in the test set.

It can be observed that, for the bottom-layer nodes, the larger the average passenger flow, the more beneficial the addition of the HP coordination module. This is mainly because the passenger flow sequences with larger volume have larger loss values and thus dominate the training process of coordinating the initial prediction results. The HP coordination module leads to an increase in prediction errors for this station. This is primarily because the initial prediction results are mapped onto the bottom-layer nodes in the HP coordination module. Then, the BU method is used to obtain the top-layer node predictions, which makes it challenging to model the top-layer node directly.
\subsubsection{Ablation Studies}
To address aspects \textbf{\hyperlink{(5)}{(5)}} and \textbf{\hyperlink{(6)}{(6)}}, we conduct the ablation experiments by individually removing the patch operation or hierarchical message passing mechanism. The defined explicitly as follows:

+IPF (w/o PA): The IPF-HMGNN without the patch operation.

+IPF (w/o HM): The IPF-HMGNN without the hierarchical message passing mechanism.

\begin{figure}[ht]
	\centering
	\includegraphics[width=14cm]{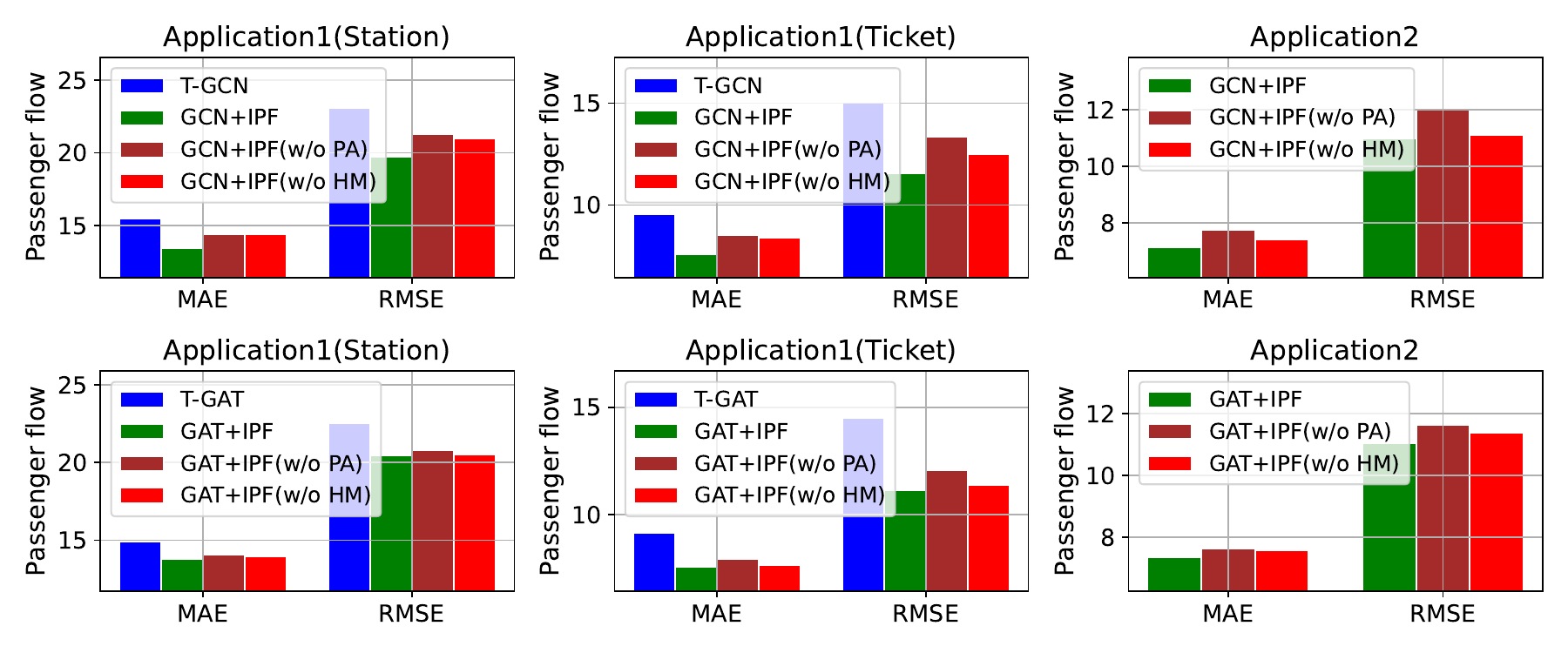}
	\caption{Ablation experiment of IPF-HMGNN.}
	\label{lu21}
\end{figure}

The results of executing prediction tasks with the above frameworks are shown in Fig.~\ref{lu21}. It can be observed that removing the patch operation or hierarchical message passing mechanism leads to a decrease in the prediction performance of the IPF-HMGNN. Additionally, the +IPF (w/o PA) and +IPF (w/o HM) outperform traditional GNN models in prediction errors, which indicates that using the patch operation and hierarchical message-passing mechanism individually in the IPF-HMGNN is effective. Finally, +IPF (w/o PA) has a more prediction error than +IPF (w/o HM), indicating that the patch operation is more critical than the hierarchical message passing operation.
\subsubsection{Application of IPF-HMGNN on existing prediction models}
Finally for the last aspect \textbf{\hyperlink{(7)}{(7)}}, we apply the proposed IPF-HMGNN to the widely used existing traffic prediction model, including the spatiotemporal graph model and multi-view graphs traffic flow prediction model, as follows:

\noindent \textbf{STGCN (Spatio-Temporal Graph Convolutional Networks) \citep{yu2017spatio}:} We use graph convolution operation in space and one-dimensional convolution in time.

\noindent \textbf{PVCGN (Physical-Virtual Collaboration Graph Network) \citep{liu2020physical}:} We merge the physical topology, similarity and correlation graphs into graph convolution-gated recurrent units.

\begin{figure}[ht]
	\centering
	\includegraphics[width=7.5cm]{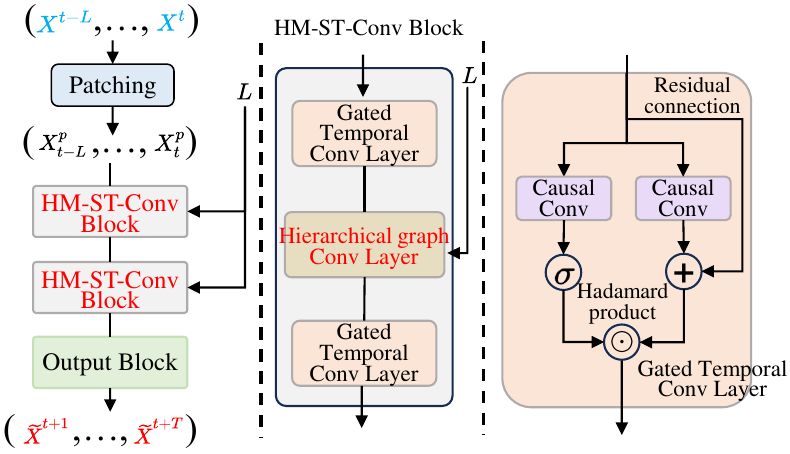}
	\caption{Application of IPF-HMGNN on STGCN.}
	\label{lu22}
\end{figure}

\begin{figure}[ht]
	\centering
	\includegraphics[width=16.2cm]{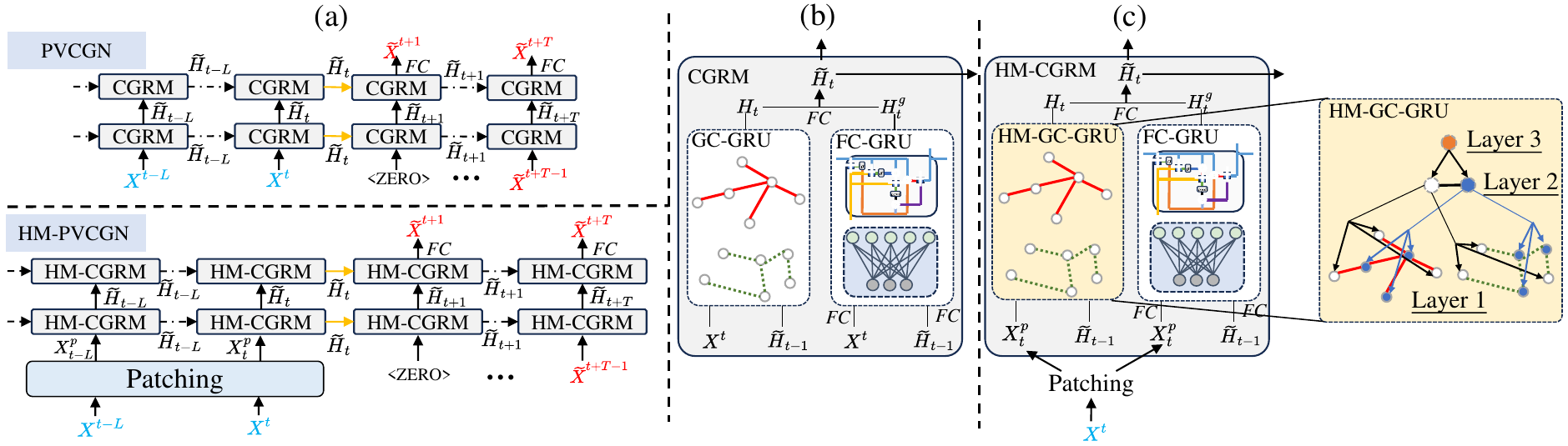}
	\caption{Application of IPF-HMGNN on PVCGN, (a) is the models of PVCGN and HM-PVCGN; (b) is the CGRM module; (c) is the HM-CGRM.}
	\label{lu23}
\end{figure}

We apply the IPF-HMGNN on the STGCN to obtain the HM-STGCN model, as shown in Fig.~\ref{lu22}. The graph convolutional layer in STGCN is replaced by a hierarchical graph convolutional, and input features are extracted through the patching operation. Similarly, we apply the IPF-HMGNN on the PVCGN to obtain the HM-PVGCN model, as shown in Fig.~\ref{lu23}. In HM-PVCGN, the hierarchical message-passing graph convolutional GRU (HM-GC-GRU) replaces the GC-GRU in PVCGN. Additionally, the patch operation is used to extract the input features.

The above models are used for the TP task. In station passenger flow prediction scenario, the physical, correlation, and similarity graphs are used in PVGCN and HM-PVCGN. In per ticket type passenger flow prediction scenario, the correlation and similarity graphs are used. The results are shown in Tables~\ref{tab5} and~\ref{tab6} which the optimal metrics are highlighted in bold black, while the second-best metrics are underlined. It can be observed that the PVCGN tends to have better performance than the STGCN model due to its ability to model the local and global evolution characteristics. Furthermore, multi-views graphs are used in PVCGN to model passenger flow sequence correlation. Finally, for the STGCN and PVCGN models, training under the IPF-HMGNN can significantly improve their performance.
\begin{table}[ht]
  \centering
  \caption{Prediction errors of different methods in station passenger flow prediction scenarios.}
    \begin{tabular}{ccccccc}
   \toprule
   Target& \multicolumn{2}{c}{Ordinary station} & \multicolumn{2}{c}{Interchange station} &\multicolumn{2}{c}{All station} \\ 
   \midrule
   Method&MAE	&RMSE	&MAE&	RMSE	&MAE	&RMSE\\
    \midrule
    STGCN&12.133 &	17.992&	61.116& 	85.146& 	14.332& 	21.342 \\
   PVCGN&\underline{9.962} 	&\underline{14.661} &	67.253 &	93.512 &	13.357 &	19.717 \\
    HM-STGCN&11.248 &	16.660 &	\underline{47.932} &\underline{64.254} &\underline{12.756} &\underline{18.965}\\ 
    HM-PVCGN&\textbf{9.448} &\textbf{13.739} 	&\textbf{43.068} &\textbf{59.906} &	\textbf{12.165} &\textbf{17.778} \\
   \bottomrule
    \end{tabular}
  \label{tab5}
\end{table}
\begin{table}[ht]
  \centering
  \caption{Prediction errors of different methods in per ticket passenger flow prediction scenarios.}
    \begin{tabular}{ccccccccc}
   \toprule
   Target& \multicolumn{2}{c}{SJT} & \multicolumn{2}{c}{CC} &\multicolumn{2}{c}{All card} &\multicolumn{2}{c}{Interchange station}\\ 
   \midrule
   Method&MAE	&RMSE	&MAE&	RMSE	&MAE	&RMSE&MAE	&RMSE\\
    \midrule
    STGCN&41.228 &	69.581 	&27.866 &	36.978& 	8.913 &	13.556 	&53.111 &	83.073 \\
    PVCGN&\underline{29.490} &\underline{49.538} 	&\underline{22.165} &\underline{31.412} &	7.889 &	12.193 &	49.850 &	74.517 \\
    HM-STGCN&36.489 &	62.054 &	24.287 	&35.450 &	\underline{6.633} &	\underline{11.073} &	\underline{42.481} &\underline{67.647} \\
    HM-PVCGN&\textbf{26.518} &\textbf{45.501} &\textbf{	21.072} &\textbf{29.675} &\textbf{6.147} &\textbf{10.068}&\textbf{41.732} &\textbf{62.628} \\
   \bottomrule
    \end{tabular}
  \label{tab6}
\end{table}
\section{Conclusion}
In this study, we proposes the IPF-HMGNN, a novel integrative prediction framework based on multi-level message-passing graph neural networks, for short-term metro passenger flow. The framework leverages the patch strategies from the natural language processing field to handle the historical passenger flow inputs. The depthwise separable convolution is used to extract local and global features. Then, the multi-level passenger flow sequences based on clustering are created, and a hierarchical message passing in GNN is developed to model the correlations between passenger flow sequences. Further, the GRU is employed to capture the temporal features of passenger flow sequences within different clusters, and the initial predictions for each sequence are obtained. Lastly, a deep learning-based hierarchical coordination model is developed to adjust the initial predictions. The key findings from the experimental results using actual data collected from the Wixu metro system network in China include:
\begin{itemize}
\item For traditional prediction tasks, the proposed IPF-HMGNN effectively improves the GNN model's fitting ability for off-peak passenger flows.

\item For hierarchical prediction tasks, the IPF-HMGNN can fulfill the hierarchical constraints and significantly reduces GNN model prediction errors.

\item Hyperparameter and robustness analyses demonstrate that the proposed IPF-HMGNN performs well in different bottom networks and GNN models.

\item Analysis of different hierarchical layer graphs support the rationale behind hierarchical structures and provides suggestions for creating hierarchies in practical applications.

\item Analysis of the hierarchical prediction module shows that the impact of various types of passenger flow sequences in the hierarchical coordination process is different. The degree of influence on bottom-layer node passenger flow sequences is related to their base passenger volume.

\item Ablation experiments demonstrate the practicality of each component in the IPF-HMGNN.

\item The IPF-HMGNN can be well integrated with other existing models (e.g. STGCN and PVCGN), demonstrating its practical application potential.
\end{itemize}

The patch idea and the multi-level modeling concept can be applied to other transportation demand prediction tasks, such as traffic flow and electric vehicle charging demand prediction. Moreover, they can also be extended to other prediction tasks, such as financial forecasting and residential electricity demand prediction. 

The main limitations of this study include the cutting of patches and the selection of the number of hierarchical layers. In practice, one can explore cutting historical passenger flow sequences into differently shaped subsequences for modeling or consider creating adaptive subsequence construction methods. Additionally, further research could involve creating more hierarchical layers based on the specific prediction task. Lastly, future research can consider other node information propagation strategies to enhance modeling capabilities and broaden application scope. 
\section*{CRediT authorship contribution statement}
\textbf{Wenbo Lu:} Conceptualization, Methodology, Software, Validation, Formal analysis, Writing – original draft, Writing – review \& editing. \textbf{Yong Zhang:} Resources, Data curation, Supervision. \textbf{Hai L.Vu:} Conceptualization,  Writing – review \& editing, Supervision. \textbf{Jinhua Xu:} Writing – review \& editing.  \textbf{Peikun Li:} Writing – review \& editing. 
\section*{Declaration of competing interest}
The authors declare that they have no known competing financial interests or personal relationships that could have appeared to influence the work reported in this paper.
\section*{Acknowledgments}
This work was supported in part by the National Natural Science Foundation of China under Grant 72071041 and the Jiangsu Province Key R\&D Program under Grant BE2021067. The first author is grateful to the China Scholarship Council (CSC) for financially supporting his visiting program at Monash University (No. 202306090270). 
\section*{Data availability}
Data will be made available on request.
\section*{\hypertarget{Appendix A} Appendix A:}
\noindent \textbf{(1) Traditional Prediction Models:}

\noindent \textbf{HA:} This model predicts the future passenger flow using the average passenger flow from the corresponding periods in historical data.

\noindent \textbf{MLP:} A type of artificial neural network with a feedforward structure that maps a set of input vectors to a set of output vectors.

\noindent \textbf{LSTM:} A type of recurrent neural network (RNN) designed to capture the long-term dependencies of historical input data. It performs well in modeling the temporal sequences of data.

\noindent \textbf{GRU:} A variant of the LSTM network with a simpler structure. It is designed to handle the long-term dependencies in the historical input sequences.

\noindent \textbf{GCN:} This model captures the local structural information between nodes through multiple layers of graph convolution operations, learning node representations.

\noindent \textbf{GAT:} Introduces the attention mechanisms, allowing nodes to assign different weights during information aggregation, enabling flexible modeling of contributions from different nodes.

\noindent \textbf{SGC:} A simplified version of the graph convolution model, omitting non-linear transformations with activation functions and parameters.

\noindent \textbf{GIN:} Captures the global structural information of the graph using a graph isomorphism network structure. It aggregates node information through multiple iterations of simple local updates.

\noindent \textbf{SAGE:} Utilizes sampling and aggregation strategies by sampling a subset of nodes from each node's neighbors and aggregating the sampled information to generate node representations.

\noindent \textbf{(2) Hierarchical Prediction Models:}

\noindent \textbf{BU:} This model predicts the passenger flow for bottom-level nodes and aggregates these predictions to obtain predictions for higher-level nodes.

\noindent \textbf{MO:} We predict the passenger flow for middle-level nodes and allocate the predictions for bottom-level nodes based on the proportion of passenger flow each bottom-level node contributes to its parent node. In addition, we sum these predictions to yield the parent node's passenger flow sequence prediction.

\noindent \textbf{TD:} We predict the passenger flow for top-level nodes and allocate predictions for bottom and middle-level nodes based on the proportions of passenger flow that each node contributes to its parent nodes.
\section*{\hypertarget{Appendix B}Appendix B:}
The parameter settings of the baseline models are as shown in Table~\ref{tab7}.
\begin{table}[ht]
  \centering
  \caption{The parameter settings of the baseline model.}
    \begin{tabular}{cccccc}
    \toprule
    Hidden node&Layer number&Dropout&Epochs&Learning rate&Activate function\\
    \midrule
    72&2&0&100&0.001&ReLU\\ 
    \bottomrule
    \end{tabular}
  \label{tab7}
\end{table}
\bibliography{inference}
\bibliographystyle{cas-model2-names}

\end{document}